# Density Dependent Hadron Field Theory

C. Fuchs[1], H. Lenske[2] and H.H. Wolter[1]

[1] Sektion Physik, Universität München,
Am Coulombwall 1, D-85748 Garching, Germany

[2] Institut für Theoretische Physik, Universität Giessen,
Heinrich-Buff-Ring 16, D-35392 Giessen, Germany

July 29, 1995

### Abstract

A fully covariant approach to a density dependent hadron field theory is presented. The relation between in–medium NN interactions and field–theoretical meson–nucleon vertices is discussed. The medium dependence of nuclear interactions is described by a functional dependence of the meson-nucleon vertices on the baryon field operators. As a consequence, the Euler-Lagrange equations lead to baryon rearrangement self–energies which are not obtained when only a parametric dependence of the vertices on the density is assumed. It is shown that the approach is energy–momentum conserving and thermodynamically consistent. Solutions of the field equations are studied in the mean–field approximation. Descriptions of the medium dependence in terms of the baryon scalar and vector density are investigated. Applications to infinite nuclear matter and finite nuclei are discussed. Density dependent coupling constants obtained from Dirac–Brueckner calculations with the Bonn NN-potentials are used. Results from Hartree calculations for energy spectra, binding energies and charge density distributions of $^{16}O$, $^{40,48}Ca$ and $^{208}Pb$ are presented. Comparisons to data strongly support the importance of rearrangement in a relativistic density dependent field theory. Most striking is the simultanuous improvement of charge radii, charge densities and binding energies. The results indicate the appearance of a new "Coester line" in the nuclear matter equation of state.

# 1 Introduction

The modern approach to nuclear structure is based on relativistic field theories describing nuclear matter as a strongly interacting system of baryons and mesons. The prototype of such a theory is QHD [1, 2, 3] in which nucleons are coupled in a minimal way to a scalar ($\sigma$) and a vector ($\omega$) meson. A variety of extensions have been studied including the isovector $\rho$ meson, electromagnetic interactions and non–linear meson self–interactions [4]. The theory is thermodynamically consistent and the covariance of the field equations is manifest [2]. The model is also applied to systems beyond normal nuclear matter, e.g. to strange matter and hypernuclei as in ref. [5].

In view of the success of the QHD models it is tempting to derive a hadron quantum field theory from a more microscopic approach to nuclear interactions. A derivation from QCD dynamics of quarks and gluons is a challenging but hitherto unsolved problem. Close to the ground state of nuclear matter confinement is the prevailing mechanism and a descripton in terms of baryons and mesons and their interactions is appropriate. But also on the hadronic level an *ab initio* calculation of the quasiparticle properties of nucleons and mesons in a nuclear medium is theoretically and numerically very involved. Except for infinite nuclear matter and a few light nuclei [6, 7] such calculations seem to be unfeasible for the near future. The advantages of a microscopic description are obvious because the empirical coupling constants could be replaced by microscopically derived values. An even more important aspect is a deeper insight into the dynamical content of effective field theories with respect to many–body dynamics in a nuclear systems. In a nuclear medium hadrons are surrounded by a cloud of polarized matter and, as a consequence, in–medium interactions differ significantly from nucleon-nucleon (NN) interactions in free space. In QHD–models the complicated many–body dynamics of strong interactions at low energy are contained effectively, but unaccessibly, in empirical meson–baryon coupling constants and non–linear meson self–interactions. The successful description of nuclear properties indicates that essential aspects of low energy strong interactions are accounted for by QHD. It is therefore reasonable to attempt a formulation which retains the basic structure of QHD but provides a more direct access to many–body dynamics. In this work we study specifically the question how to implement many–body effects in nuclear interactions into a hadron quantum field theory. A first account of an relativistically invariant approach to a density dependent (DD) hadron field theory was given in ref.[8]. Here, the theoretical background and applications to infinite matter and finite nuclei are discussed in more detail.

A widely used and successful description of in–medium interactions is given by Brueckner theory. The screening of NN–interactions in a nuclear medium is described non-perturbatively by an explicit treatment of two–body correlations [9, 10]. Over the last years, extensions to relativistic Dirac–Brueckner (DB) theory have been investigated by several groups, e.g. refs. [11, 12, 13, 14, 15, 16, 17, 18]. But again, from a practical point of view an efficient and reliable approximation scheme is required before applications to finite nuclei become feasible. A tractable approach for finite nuclei is the local density approximation (LDA) which originally was introduced in non–relativistic Brueckner theory [19]. The essential step of the LDA is to include many–body correlations into an effective two–body interaction rather than to treat them explicitly on the level of wave functions. This leads to the Brueckner G–matrix which retains the boson exchange picture of nuclear interactions and accounts for medium effects by density dependent interaction strengths.

An important result of relativistic theory is that the bulk of the in–medium screening is accounted for by a dependence of DB self–energies on the local baryon number density while the momentum dependence of interactions in the (positive energy) Fermi–sea is rather weak [13, 15, 20]. This offers the possibility to approximate relativistic many–body dynamics by a DD hadronic field theory. In practice, a two–body G–matrix is calculated first in nuclear matter for a fixed baryon number density and then self–energies of baryons are derived [16, 21]. Because of the smooth dependence on density the DB self–energies are expressed by self–energies of a structure as in the $\sigma$–$\omega$ model. A set of meson fields with fixed masses is chosen and the meson–baryon coupling constants are adjusted to the DB self–energies.

It is common practice to use the infinite matter DB coupling constants in an effective hadron quantum field theory. A QHD–type Lagrangian is chosen and the screening of interactions is taken into account by effective meson–nucleon vertices depending on the local baryon number density [17, 21, 22, 23]. By construction the DB self–energies and the total energy density in infinite matter are reproduced. Because the structure of such a density dependent (DD) hadron field theory is much simpler than the original DB calculations applications to finite nuclei become possible.

However, this relativistic LDA leads to a field theory of an unsatisfactory structure as was pointed out in ref. [8]. Part of the problems become apparent when the theory is applied to an inhomogenuous system like a finite nucleus. Then one finds immediately that a parametric dependence of vertices on the local number density transforms into a dependence of the coupling constants on space–time coordinates. The Lagrangian ceases to be a Lorentz–scalar and questions on causality, the covariance of the field equations and the thermodynamical consistency of the theory arise. If at all, such an approach provides an effective mean–field Lagrangian which is valid in the nuclear rest frame only.

In order to obtain an intrinsically consistent field theory of wider applicability the Lagrangian must be formulated as a functional of field amplitudes only. Approximations should be introduced in later stages of the calculations after the field equations have been derived. These are exactly the steps by which the mean–field approximation to the full QHD Lagrangian is obtained [1, 2]. The relation between an explicit many–body description of in–medium interactions and a field theory with effective density dependent meson–nucleon vertices is by no means obvious. The two–body NN amplitude does not enter directly into a field–theoretical formulation. In a Lagrangian nuclear interactions are described by a vertex to which baryon and meson fields are coupled. The vertices are determined by "amputated" diagrams which are obtained by cutting the two–body amplitude and a resummation of interactions. In section 2.1 the diagrammatic structure of meson–nucleon vertices and their use in an effective field theory are briefly discussed. Vertex caculations are a standard problem of quantum field theory [24]. The discussion refers mainly to former works on the electron-photon vertex in QED [25] and electron–phonon vertices in solid state physics [26]. The many–body theoretical methods developed in these fields are especially useful for investigations of medium effects in meson–baryon vertices. In principle, the method allows to calculate the in–medium vertices without the necessity of a numerical fit to nuclear matter T-matrices or self–energies. In this work, however, such an extended many–body calculation is not considered. We rather assume that the nuclear matter vertices are known from other sources like DB calculations. A Lorentz–invariant functional of the baryon field operators is defined to project the nuclear matter results onto the meson–nucleon vertices of an effective density dependent field theory. In sect. 2.2 the model Lagrangian is introduced. Meson and baryon field equations are derived in sect. 2.3. The Euler–Lagrange equations lead to additional baryon rearrangement self–energies from the variation of the vertices. In sections 2.4 and 2.5 field equations and rearrangement self–energies are studied for a description of medium effects by the scalar and vector baryon densities, respectively. In the latter case, covariance requires to use the square of the four–vector current rather than simply the time–like component only.

The baryon rearrangement self–energies are not obtained in the conventional formulation of a density dependent field theory because the vertices are assumed to depend only parametrically on the local density which is obtained at the end of the calculation. Commonly, a self–consistent procedure is used and the vertices are readjusted in each iteration. Such an approach accounts for the dependence of interactions on the bulk variations of the background medium. The dynamical adjustment of vertices steming from the polarization of the many–body background by a baryon, however, is neglected. From non–relativistic Hartree–Fock (HF) theory it is known that rearrangement accounts for an important part of the nuclear mean–field and is indispensible for a good description of nuclear binding energies and density distributions [27]. From non–relativistic structure theory it is also known that rearrangement includes ground state correlations from high momentum excitations of the background medium into the nuclear mean–field [9, 29].

In section 3 solutions of the field equations are studied in the mean–field or Hartree limit. In Hartree approximation the otherwise highly non–linear field equa-tions are reduced to a tractable form because the vertex functionals can be replaced by functions of expectation values. In the mean–field limit field equations of a structure similar to conventional DD Hartree theory are obtained but with inclusion of rearrangement. In sect. 3.1 the theory is applied to infinite nuclear matter. Energy–momentum conservation and the thermodynamical consistency of the model are shown. An important result is the cancellation of rearrangement energies in the energy density.

In section 4 the theory is used to describe ground state properties of the doubly closed shell nuclei $^{16}O$,$^{40,48}Ca$ and $^{208}Pb$. Details of the numerical calculations are discussed in section 4.1. Once the density dependence of the coupling constants is known the numerical effort to solve the Hartree field equations self–consistently is comparable to QHD. Density dependent $\sigma$ and $\omega$ meson coupling constants are used which were derived from infinite matter DB calculations using the Bonn A,B and C NN potentials [15, 22, 30, 31]. The main difference between the three potentials is a different strength of the tensor interaction. The coupling constants already include exchange because they were obtained from fits to DB self–energies. Therefore, in order to avoid overcounting they are used in Hartree calculations only. A Hartree–Fock calculation would require to parameterize the NN T–matrix elements directly as in refs. [16, 23]. In sections 4.2 and 4.3 results of density dependent relativistic Hartree calculations for energy spectra and density distributions are compared to data. The best description is obtained with the Bonn A parameter set and including rearrangement. With rearrangement the description of measured single particle spectra and charge densities is significantly improved. The importance of rearrangement in finite nuclei is seen very clearly also from the dependence of the binding energy per particle on the central density. Experimental values for nuclei between $^{16}O$ and $^{208}Pb$ form a narrow band which can be considered to represent an equation of state for finite nuclei. The theoretical results for the three NN-potentials arrange on "Coester lines" and the data are reproduced only when rearrangement is included. The paper closes in sect. 5 with a summary and concluding remarks.

# 2 Density Dependent Hadron Field Theory

## 2.1 From Nuclear Matter to Finite Nuclei

Once a hadronic Lagrangian has been defined medium effects can be treated systematically with many–body theory. For the sake of a tractable model, however, an approximate treatment of medium effects is necessary. Non–relativistically, the medium dependence of NN interactions is well described by energy functional methods [29, 32]. Relativistically, a Lagrangian formulation with baryons and mesons has to be used and medium effects are described by effective density dependent

meson–baryon vertices. A relation between microscopic and effective meson–nucleon vertices is, in principle, given in the local density approximation to DB self–energies [20, 21, 23]. The use of nuclear matter DB results in an effective hadron field theory, however, deserves a closer discussion because the relation of the two approaches is not obvious. A number of non–trivial approximations are necessary before the link to a field theory of mesons and baryons is obtained. At the end, the theory should account reliably for the complexity of a DB calculation by a limited number of parameters. In this section the most important steps in going from a nuclear matter T-matrix to a field theory with medium dependent meson–baryon vertices are outlined.

As a matter of fact, the two–body DB amplitudes do not enter directly into a field–theoretical description. Rather, effective medium dependent meson–baryon vertices have to be extracted from the DB T-matrix $\mathcal{T}$. As a first step a parameterization of $\mathcal{T}$ in terms of effective meson propagators, e.g. as in ref.[23], must be derived. For the following discussion we assume this to be given. Then, we can proceed as in refs. [25, 26] and consider directly the vertices describing the coupling of mesons and baryons. In the usual formulation $\mathcal{T}$ is obtained by an integral equation from the free–space NN boson exchange interactions $V$ [13, 23]. For our purpose it is more convenient to proceed differently. Let $\mathcal{G}$ denote the two point baryon–propagator including the full in–medium self–energy $\Sigma$. The effective two–particle propagator is given by $\mathcal{G}^{12} = i\mathcal{G}\mathcal{G}$ and $\mathcal{G}^{12}_F = \mathcal{Q}_F i\mathcal{G}\mathcal{G}\mathcal{Q}_F$ is the projection of $\mathcal{G}^{12}$ onto the Fermi–sphere of occupied states. In a symbolic notation $\mathcal{T}$ is expressed by the free space NN T-matrix $T$

$$\mathcal{T} = T - T\mathcal{G}^{12}_F \mathcal{T} \quad ; \quad T = V + V\mathcal{G}^{12}T \qquad (1)$$

and the main difference to the usual approach is to disregard blocking first and then restore the Pauli–principle by subtracting the contributions from the interior of the Fermi–sphere. For our purposes Eq.(1) has the advantage that the medium projector $\mathcal{Q}_F$ appears explicitly instead of the complementary projector $\mathcal{Q} = 1 - \mathcal{Q}_F$. Also, $T$ is free of the "hard core" singularities contained in $V$ and thus easier to handle. As a rather schematic *ansatz* (see e.g. ref.[23] for a realistic description) we assume that

$$\mathcal{T}(1, 2) \simeq \Gamma(1)\Delta(1, 2)\Gamma(2) \quad ; \quad T(1, 2) \simeq \gamma(1)\Delta(1, 2)\gamma(2) \qquad (2)$$

are given by an effective meson propagator $\Delta$ and the in–medium and free vertices $\Gamma$ and $\gamma$, respectively. The vertices $\gamma = g_0 e_D$ include the free meson–baryon coupling strength $g_0 = g_\sigma, g_\omega \ldots$ and the Dirac–structure is taken into account by $e_D = \mathbb{1}, \gamma_\mu \ldots$. Using the same propagators in $\mathcal{T}$ and $T$ corresponds to assume that the mesons are unaffected by the medium. Following ref.[25] the free and in–medium meson–baryon vertices are related by a Bethe–Salpeter type equation

$$\Gamma = \gamma - \Gamma\mathcal{G}^{12}_F T \quad . \qquad (3)$$

Note, that only states from inside the Fermi–sphere appear as intermediate states because of $\mathcal{G}_F^{12}$. From the diagrammatic structure, Fig.1, it is seen that $\Gamma$ includes the full ladder series of repeated actions of $T$ between the in– and out–going baryons. Eq.(2) and Fig.1 express an important relation: The one boson exchange parameterization of $\mathcal{T}$ corresponds to a resummation of interactions such that the ladder series is effectively shifted to the vertices. Because of Eq.(1) the ladder series is now defined in terms of $T$ rather than by the bare NN potential $V$.

The weak momentum dependence of self–energies [20] indicates that the vertices themselves are only weakly dependent on momentum. It is therefore reasonable to define effective vertices which depend only on the density of nuclear matter by averaging Eq.(3) over the Fermi–sphere

$$\Gamma(k_F) \;=\; \frac{tr\left[\Gamma\mathcal{G}_F\right]}{tr\left[e_D\mathcal{G}_F\right]} \qquad (4)$$

which means to close the baryon legs in Fig.1. $\mathcal{G}_F = \mathcal{G}\mathcal{Q}_F$ is the nuclear matter two–point function. In nuclear matter the vertex can be written as $\Gamma_{nm} = e_D\Gamma(k_F)$ where the dependence on the Fermi momentum $k_F$, or equivalently on the density $\rho_{nm} = 2k_F^3/(3\pi^2)$, is introduced by $\mathcal{G}_F$. The full nuclear matter self–energy is obtained as in [2, 13]

$$\Sigma = tr[\mathcal{T}\mathcal{G}_F] = \Gamma tr[\Delta\Gamma\mathcal{G}_F] \quad . \qquad (5)$$

Exchange is assumed to be included, e.g. by a Fierz–transformation of $\mathcal{T}$ [24, 28]. The diagrammatic structure of $\Sigma$ is shown in Fig.2. Eq.(5) implies to define the in–medium meson field

$$\Phi = tr\left[\Delta\Gamma\overline{\Psi}\Psi\right] = tr\left[\Delta\gamma\mathcal{G}_F\right]\Gamma(k_F) \qquad (6)$$

and to use the vertex $\mathcal{L}_{int} = \Phi\Gamma\overline{\Psi}\Psi$ in a Lagrangian which also contains a kinetic and mass term for the meson field $\Phi$. The latter can be deduced from $\Delta^{-1}$. In nuclear matter the parameterized vertex, Eq.(4), can safely be used in the Lagrangian because $k_F$ enters as an external parameter. Inserting Eq.(4) into Eq.(5) the DB self–energies are seen to be reproduced by construction.

A different situation is encountered in a finite nucleus. Then the density becomes a dynamical quantity which has to be determined self–consistently by an appropriate adjustment of a chemical potential. In ref. [8] it was shown that an intrinsically consistent field theory is obtained when the meson–baryon vertices are chosen as functionals of the baryon field operators. Here, a more general approach is discussed by which the mapping from the nuclear matter DB results to a field theoretical formulation is easily obtained. Suppose that $\Gamma(\rho_{nm})$ was calculated in infinite matter for a sufficiently wide range of densities $\rho_{nm}$. Using the baryon current operator $\hat{j}^\mu = \overline{\Psi}\gamma^\mu\Psi$ we define the functional

$$\Gamma(\hat{j}^2) = \int_0^\infty \Gamma(\rho_{nm})\delta\left(\rho_{nm}^2 - \hat{j}_\mu\hat{j}^\mu\right) 2\rho_{nm}d\rho_{nm} \qquad (7)$$

which provides the mapping of the infinite matter results onto a field–theoretical formulation. The four–vector $\hat{j}^\mu$ is determined in the nuclear rest frame. For infinite matter or a spherical nucleus, respectively, the expectation value is $\langle 0|\hat{j}^\mu|0\rangle = \delta^{\mu 0}\rho_B$ where $\rho_B$ is the baryon number density. Hence, the functional is Lorentz–invariant and depends only on field amplitudes. Assuming that a particular infinite matter configuration is determined unambiguously by the value of the number density $\rho_{nm}$ the integral provides a point–wise one–to–one mapping from the infinite matter vertices $\Gamma(\rho_{nm})$ to a field–theoretical meson–nucleon vertex $\Gamma(\hat{j}^2)$. An extension to multivariant parameterizations is easily obtained by augmenting the required numbers of projections, e.g. for asymmetric matter in terms of the isoscalar and the isovector baryon densities. For more general theories of $\mathcal{T}$ beyond the DB ladder approximation a description in terms of n–body baryon and meson operators, respectively, could be found necessary [8].

## 2.2 The Model Lagrangian

The model Lagrangian [8] includes baryon fields $\Psi$ and the isoscalar $\sigma$ and $\omega$ mesons, the isovector $\rho$ meson and the photon ($\gamma$),

$$\begin{aligned}
\mathcal{L} &= \mathcal{L}_B + \mathcal{L}_M + \mathcal{L}_{int} \\
\mathcal{L}_B &= \overline{\Psi}(i\gamma_\mu \partial^\mu - M)\Psi \\
\mathcal{L}_M &= \frac{1}{2}(\partial_\mu \Phi \partial^\mu \Phi - m_s^2 \Phi^2) - \frac{1}{2}\sum_{\kappa=\omega,\rho,\gamma}\left(\frac{1}{2}F^{(\kappa)}_{\mu\nu}F^{(\kappa)\mu\nu} - m_\kappa^2 V^{(\kappa)}_\mu V^{(\kappa)\mu}\right) \\
\mathcal{L}_{int} &= \overline{\Psi}\Psi\Gamma_\sigma(\hat{\rho}_0)\Phi - \overline{\Psi}\gamma_\mu\Psi\Gamma_\omega(\hat{\rho}_0)V^\mu_\omega - \overline{\Psi}\ \gamma_\mu\Psi g_\rho \mathbf{V}^\mu_\rho - \overline{\Psi}\frac{1}{2}(1+\tau_3)\gamma_\mu\Psi e A^\mu
\end{aligned} \quad (8)$$

where $\mathcal{L}_B$ and $\mathcal{L}_M$ denote the Lagrangians of free baryons and mesons, respectively, and their interactions are described by $\mathcal{L}_{int}$.

$$F^{(\kappa)}_{\mu\nu} = \partial_\mu V^{(\kappa)}_\nu - \partial_\nu V^{(\kappa)}_\mu \quad (9)$$

is the field strength tensor for a vector meson ($\kappa = \omega,\rho$) or the photon ($\kappa = \gamma$). The Lagrangian resembles those of the $\sigma$–$\omega$ model [1, 2] and the DD field theories of e.g. refs.[12, 15, 16, 14, 21]. The important difference lies in our treatment of medium effects in the meson–baryon vertices.

For a more transparent presentation only the $\sigma$ and $\omega$ vertices $\Gamma_\sigma$ and $\Gamma_\omega$, respectively, are taken to be density dependent but extensions to other meson channels are obvious. Also, in the applications vertices from DB calculations in symmetric nuclear matter will be used which naturally provide information on the isoscalar mesons only. The vertices are assumed as in Eq.(7) but for a more general formulation a Lorentz-scalar functional $\hat{\rho}_0 = \hat{\rho}_0(\overline{\Psi},\Psi)$ is used whose form is specified later. It is assumed that the nuclear matter vertices entering into Eq.(7) are given in terms of the expectation value $\rho_{nm} = \langle nm|\hat{\rho}_0|nm\rangle$. In order to retain commutator relations of Dirac operators and vertices $\hat{\rho}_0$ must contain an even number of baryon field operators. An obvious choice is to use $\hat{\rho}_0 = \overline{\Psi}\Psi$ leading to a scalar density dependence (SDD). The connection to conventional parameterizations of DB–vertices is obtained with the vector density dependence (VDD) $\hat{\rho}_0^2 = \hat{j}_\mu \hat{j}^\mu$ as discussed before. These two cases will be investigated in more detail in sections 2.4 and 2.5, respectively.

It is worthwhile to emphasize that other choices are possible as well. In fact, the consistency of the theory is preserved for any Lorentz–invariant combination of baryon and meson field operators. However, as shown below the dynamics of the fields are directly affected by the structure of $\hat{\rho}_0$. This leads to further constraints on $\hat{\rho}_0$ by physical reasons. The description of medium effects by a functional of baryon field operators only, for example, leaves the meson field equations unchanged and ascribes many–body effects completely to the baryon self–energies. From a conceptual point of view this has the advantage that only the baryon sector of the model is affected. Also, such a description is strongly supported by DB results and non–relativistic many–body calculations [9, 29]. From a more practical point of view uncertainties on parameters are kept on a controllable level and the relation to QHD–type approaches is more transparent.

## 2.3 The Field Equations

From the Euler-Lagrange equations meson field equations are obtained which in form resemble those of QHD [1, 2]. The differences reside in the source terms of the $\sigma$ and $\omega$ fields which include in–medium correlations through the vertices:

$$(\partial_\mu \partial^\mu + m_\sigma^2)\Phi = \Gamma_\sigma(\hat{\rho}_0)\overline{\Psi}\Psi \tag{10}$$

$$(\partial_\nu F^{(\omega)\mu\nu} + m_\omega^2)V_\omega^\mu = \Gamma_\omega(\hat{\rho}_0)\overline{\Psi}\gamma^\mu\Psi \tag{11}$$

$$(\partial_\nu \mathbf{F}^{(\rho)\mu\nu} + m_\rho^2)\mathbf{V}_\rho^\mu = g_\rho \frac{1}{2}\overline{\Psi}\ \gamma^\mu\Psi \tag{12}$$

$$\partial_\nu F^{(\gamma)\mu\nu} = e\frac{1}{2}\overline{\Psi}(1+\tau_3)\gamma^\mu\Psi \tag{13}$$

where $\tau_3 = \pm 1$ for protons and neutrons, respectively.

An important difference to QHD and conventional formulations of DD theories appears in the baryon field equations. From Eq.(7) it is evident that the variation with respect to $\overline{\Psi}$ will also act on the vertices and thus introduce additional self–energies. Formally, this is taken into account by treating $\hat{\rho}_0$ as an additional degree of freedom which can be thought to act as an external source of many–body correlations. The variational derivative of $\mathcal{L}_{int}$, Eq.(8), is written as

$$\frac{\delta \mathcal{L}_{int}}{\delta \overline{\Psi}} = \frac{\partial \mathcal{L}_{int}}{\partial \overline{\Psi}} + \frac{\partial \mathcal{L}_{int}}{\partial \hat{\rho}_0}\frac{\delta \hat{\rho}_0}{\delta \overline{\Psi}} \quad . \tag{14}$$

$\hat{\rho}_0$ is assumed to depend on the baryon operators only,

$$\delta\hat{\rho}_0 = \frac{\partial\hat{\rho}_0}{\partial\overline{\Psi}}\delta\overline{\Psi} \quad . \tag{15}$$

By definition, the derivative of $\hat{\rho}_0$ must be proportional to $\Psi$ and covariance requires that the proportionality factor is composed of Lorentz–invariants. Here, we consider only Lorentz scalar and vector terms,

$$\frac{\partial\hat{\rho}_0}{\partial\overline{\Psi}} = [A_s + B_\nu\gamma^\nu]\Psi \tag{16}$$

with coefficients $A_s$ and $B_\nu$, respectively. With

$$S^{(r)} = \frac{\partial\mathcal{L}_{int}}{\partial\hat{\rho}_0} = \left[\overline{\Psi}\Psi\frac{\partial\Gamma_\sigma}{\partial\hat{\rho}_0}\Phi - V_\omega^\nu\hat{\jmath}_\nu\frac{\partial\Gamma_\omega}{\partial\hat{\rho}_0}\right] \tag{17}$$

the second term in Eq.(14) leads to the rearrangement self–energy

$$\Sigma^{(r)} = \Sigma_s^{(r)} + \Sigma_\nu^{(r)}\gamma^\nu \quad . \tag{18}$$

The scalar and vector parts are given by

$$\Sigma_s^{(r)} = S^{(r)}A_s \quad ; \quad \Sigma_\mu^{(r)} = -S^{(r)}B_\mu \tag{19}$$

and with the usual self–energies [1, 2]

$$\Sigma_s^{(0)} = \Gamma_\sigma(\hat{\rho}_0)\Phi \tag{20}$$

$$\Sigma^{(0)\mu} = \Gamma_\omega(\hat{\rho}_0)V_\omega^\mu + g_\rho \cdot \mathbf{V}_\rho^\mu + \frac{1}{2}(1+\tau_3)eA^\mu \tag{21}$$

the total baryon self–energies are finally obtained as

$$\Sigma_s = \Sigma_s^{(0)} + \Sigma_s^{(r)} \quad , \quad \Sigma^\mu = \Sigma^{(0)\mu} + \Sigma^{(r)\mu} \quad . \tag{22}$$

In structure the baryon field equations are unchanged

$$[\gamma_\mu(i\partial^\mu - \Sigma^\mu) - (M - \Sigma_s)]\Psi = 0 \tag{23}$$

but the dynamics are modified by the rearrangement contributions. Eqs.(16) to (19) are the central results of this section. They show that a covariant formulation of a DD hadron field theory leads naturally to rearrangement contributions. Medium effects affect the field dynamics in two different ways, namely mesons and baryons through the intrinsic density dependence of the $\Gamma_{\sigma,\omega}$ vertices and, in addition, the baryon fields by the rearrangement self–energies. The form and Lorentz structure of the latter contributions depends sensitively on the form of $\hat{\rho}_0$. In the following, two physically reasonable choices are discussed.

## 2.4 Vector Density Dependence (VDD)

In the VDD description the square of baryon vector current, $\hat{\rho}_0^2 = \hat{j}_\mu \hat{j}^\mu$, is used. The $\sigma$ and $\omega$ vertices are defined as in Eq.(7),

$$\Gamma_{\sigma,\omega}(\hat{\rho}_0) = \int_0^\infty \Gamma_{\sigma,\omega}(\rho_{nm}) \delta\left(\rho_{nm}^2 - \hat{\rho}_0^2\right) 2\rho_{nm} d\rho_{nm} \quad . \tag{24}$$

It is obvious that Lorentz–invariance would be badly violated if only the time–like component of $\hat{j}^\mu$ had be taken at this point. The link to the usual LDA description is obtained after the functional mapping which projects the nuclear matter results onto the invariant local baryon density $\hat{\rho}_0$. From the properties of the Dirac $\delta$–function one finds immediately

$$\frac{\delta \Gamma_{\sigma,\omega}}{\delta \hat{\rho}_0} = \int_0^\infty \frac{\partial \Gamma_{\sigma,\omega}(\rho_{nm})}{\partial \rho_{nm}} \delta\left(\rho_{nm}^2 - \hat{\rho}_0^2\right) 2\rho_{nm} d\rho_{nm} \tag{25}$$

such that the derivative of the nuclear matter vertex is mapped onto it's value at the local invariant number density.

Variation of $\hat{\rho}_0$ with respect to $\overline{\Psi}$ leads to

$$\frac{\delta \hat{\rho}_0}{\delta \overline{\Psi}} = \frac{\partial \hat{\rho}_0}{\partial \overline{\Psi}} = \gamma_\mu \hat{u}^\mu \Psi \tag{26}$$

where $\hat{\rho}_0 \hat{u}^\mu = \hat{j}^\mu$ with $\hat{u}^2 = \mathbb{1}$. $B_\mu = \hat{u}_\mu$ is a four–velocity and $A_s = 0$. In this case, only a vector rearrangement self–energy is obtained

$$\Sigma^{(r)\mu} = \left(\frac{\partial \Gamma_\omega}{\partial \hat{\rho}_0} V_\omega^\nu \hat{j}_\nu - \frac{\partial \Gamma_\sigma}{\partial \hat{\rho}_0} \overline{\Psi} \Psi \Phi\right) \hat{u}^\mu \tag{27}$$

which includes contributions from both the scalar and the vector fields.

## 2.5 Scalar Density Dependence (SDD)

In leading order the scalar and vector parts of $\mathcal{T}$ are determined by vertices of corresponding Lorentz structures. Thus, it is of interest to investigate a model where medium effects in $\Gamma_\sigma$ are described by a scalar density dependence $\hat{\rho}_s = \overline{\Psi}\Psi$ and the vector dependence is retained for $\Gamma_\omega$. In this case, the scalar vertex functional is chosen as

$$\Gamma_\sigma(\hat{\rho}_s) = \int_0^\infty \Gamma_\sigma(\rho_{s_{nm}}) \delta(\rho_{s_{nm}} - \hat{\rho}_s) d\rho_{s_{nm}} \tag{28}$$

where $\rho_{s_{nm}}$ is the scalar density in nuclear matter. Accordingly, the second term of Eq.(14) now includes a summation over the independent scalar and vector parameterizations. $\Gamma_\omega$ is given as in Eq.(24). $A_s = \mathbb{1}$ is non–vanishing and $B_\mu$ remains the same as in Eq.(27). Different to the VDD case, both scalar and vector rearrangement self–energies are obtained

$$\Sigma_s^{(r)} = \frac{\partial \Gamma_\sigma}{\partial \hat{\rho}_s} \hat{\rho}_s \Phi \quad , \quad \Sigma^{(r)\mu} = \frac{\partial \Gamma_\omega}{\partial \hat{\rho}_0} V_\omega^\nu \hat{j}_\nu \hat{u}^\mu \quad . \tag{29}$$

$\Sigma_s^{(r)}$ introduces an additional density dependence into the effective baryon mass. A comparison of Eq.(29) and Eq.(27) shows that also the vector rearrangement self–energy is changed. As discussed below, the differences in SDD and VDD dynamics lead to different results for the level structure of finite nuclei which allows to compare the two approaches on an empirical basis.

As a theoretically interesting side aspect the SDD *ansatz* allows to express the full $\sigma$ and $\omega$ self-energies as derivatives with respect to the densities

$$\Sigma^\mu = \Sigma^{(0)\mu} + \Sigma^{(r)\mu} = V_\omega^\nu \frac{\partial}{\partial \hat{j}^\mu} \left( \hat{j}_\nu \Gamma_\omega(\hat{j}^2) \right) \tag{30}$$

$$\Sigma_s = \Sigma_s^{(0)} + \Sigma_s^{(r)} = \Phi \frac{\partial}{\partial \hat{\rho}_s} \left( \hat{\rho}_s \Gamma_\sigma(\hat{\rho}_s) \right) \tag{31}$$

where rearrangement is included. The SDD description is easily extended to other mesons. As a rule, the vertices have to be parameterized by the densities which are the sources of corresponding meson fields and therefore are of the same Lorentz and isospin structure. For example, the $\rho$-meson would have to be treated similar to the $\omega$–meson but replacing in Eq.(24) the vector current $\hat{j}_\mu$ by the isovector current $\mathbf{j}_\mu = \overline{\Psi} \, \gamma_\mu \Psi$.

## 3 Mean–Field Theory

### 3.1 Mean–Field Interactions in Hartree Approximation

From the preceding discussion it is apparent that the field–theoretical formulation in principle includes a wider class of diagrams than considered in a DB calculation. The vertices defined in Eq.(7) contain for example contributions from the Dirac sea as well as from the full range of positive energy states as seen when the baryon field operators are inserted in quantized form. By physical arguments and because of numerical reasons DB calculations neglect vacuum contributions and ascribe medium effects solely to the polarization of the Fermi sea of positive energy valence particles. Thus, further approximations are required before DB vertices can be used in a field–theoretical approach. Here we consider a density dependent mean–field description of finite nuclei in Hartree approximation. As discussed in great detail by Boersma and Malfliet [16] this corresponds using DD coupling constants fitted to DB self–energies rather than to the in–medium NN T-matrix. As is common practice in mean–field and also DB calculations we neglect contributions from the Dirac–sea. On a more formal level this may be expressed by an explicit subtraction of vacuum expectation values [2, 24]. Here, we assume that products of fermion operators are normal ordered with respect to the Hartree ground state $|0\rangle$. The ground state is a single slater determinant of occupied fermion states. The baryons are moving in a static mean–field generated by stationary classical meson fields. Expectation values with respect to the Hartree ground state will be abbreviated e.g. as $\langle \Gamma_\sigma \rangle = \langle 0|\Gamma_\sigma|0\rangle$.

The Hartree approach allows a particularily simple treatment of the vertex functionals $\Gamma(\hat{\rho}_0)$, sect. 2.2. Using Wick's theorem,

$$\hat{\rho}_0 = \rho_0 + C(\hat{\rho}_0) \tag{32}$$

is expressed by the pure c–number valued function $\rho_0 = \langle\hat{\rho}_0\rangle$ and a remaining operator part with $\langle C \rangle = 0$. From the operator structure of $\Psi$ [2, 28] the correlation function $C$ is seen to include the normal ordered parts of $\hat{\rho}_0$ and non–stationary particle–hole type components. As a consequence expectation values of powers of $\hat{\rho}_0$ are reduced to powers of the expectation values, e.g. $\langle \hat{\rho}_0^2 \rangle = \rho_0^2$. Inserting Eq.(32) into Eq.(7) and expanding the Dirac $\delta$–function under the integral into a Taylor series the vertex functionals become

$$\Gamma_{\sigma,\omega} = \Gamma_{\sigma,\omega}(\rho_0) + C(\hat{\rho}_0)\frac{\partial}{\partial \rho_0}\Gamma_{\sigma,\omega}(\rho_0) + \ldots \quad . \tag{33}$$

In Hartree approximation the vertex functionals are therefore replaced by functions of expectation values [8]

$$\langle \Gamma_{\sigma,\omega}(\hat{\rho}_0) \rangle = \Gamma_{\sigma,\omega}(\rho_0) \tag{34}$$

which brings the originally highly non–linear field equations into a tractable form. From Eq.(33) it is seen that the rearrangement contributions are obtained from the variation of the normal ordered parts of $\hat{\rho}_0$. The Hartree meson field equations are obtained by taking expectation values on both sides of Eqs.(10) to (13). Here, we concentrate on the density dependent scalar and vector fields. The meson fields are decomposed into stationary and time dependent fluctuating parts of vanishing ground state expectation value [2], e.g for the scalar field

$$\Phi(t,\mathbf{r}) = \langle\Phi\rangle(\mathbf{r}) + \delta\Phi(t,\mathbf{r}) \quad . \tag{35}$$

Using Eqs.(33) and (32) the vertices and baryon sources are replaced by expectation values. The $\sigma$ and the $\omega$ mean–field equations become

$$(-\Delta + m_\sigma^2)\Phi = \Gamma_\sigma(\rho_0)\rho_s \tag{36}$$
$$(-\Delta + m_\omega^2)V_\omega^\mu = \Gamma_\omega(\rho_0)j^\mu \tag{37}$$

where $\Phi$ and $V_\omega^\mu$ now denote static classical fields. With the scalar and vector propagators $D_\sigma$ and $D_\omega^{\mu\nu}$ [2, 28], respectively, the solution of Eqs.(36) and (37) are

$$\Phi(\mathbf{r}) = \int d\mathbf{r}' D_\sigma(\mathbf{r},\mathbf{r}') \Gamma_\sigma(\rho_0(\mathbf{r}')) \rho_s(\mathbf{r}') \tag{38}$$

$$V_\omega^\mu(\mathbf{r}) = \int d\mathbf{r}' D_\omega^{\mu\nu}(\mathbf{r},\mathbf{r}') \Gamma_\omega(\rho_0(\mathbf{r}')) j_\nu(\mathbf{r}') \ . \tag{39}$$

The $\rho$ meson and the static Coulomb mean–field are the same as in QHD [2]. They are obtained accordingly by inserting the appropriate propagators and source terms [8]. The non–rearrangement parts of the baryon Hartree self–energies are given by

$$\Sigma_s^{(0)} = \Gamma_\sigma(\rho_0)\Phi \tag{40}$$

$$\Sigma^{(0)\mu} = \Gamma_\omega(\rho_0)V_\omega^\mu + g_\rho \cdot \mathbf{V}_\rho^\mu + \frac{1}{2}(1+\tau_3)eA^\mu \tag{41}$$

which differ from Eqs.(20) and (21) by the replacement of $\hat{\rho}_0$ by the expectaton value $\rho_0$. In the VDD Hartree description where

$$\rho_0^2 = \langle j_\mu \rangle \langle j^\mu \rangle = \rho_B^2 \tag{42}$$

both the $\sigma$ and $\omega$ vertices become functions of the local baryon density. In the SDD parameterization $\Gamma_\sigma$ is a function of the scalar density $\rho_s = \langle \overline{\Psi}\Psi \rangle$ while for $\Gamma_\omega$ the vector density dependence is retained.

The Hartree rearrangement self–energies are derived by the same approximations as above. Meson field operators are replaced by the classical fields and only the fully contracted parts of products of baryon operators are considered. As a result, the VDD rearrangement self-energy, Eq.(27), simplifies in the Hartree limit to

$$\Sigma^{(r)\mu} = \left(\frac{\partial \Gamma_\omega}{\partial \rho_0} V_\omega^\nu j_\nu - \frac{\partial \Gamma_\sigma}{\partial \rho_0} \rho_s \Phi\right) u^\mu \ . \tag{43}$$

In the nuclear restframe $u^\mu = (1,\mathbf{0})$ and

$$j_\nu V_\omega^\nu = \rho_B V_\omega^0|_{restframe} \ . \tag{44}$$

Correspondingly, the SDD Hartree scalar rearrangement self–energy

$$\Sigma_s^{(r)} = \Phi \rho_s \frac{\partial \Gamma_\sigma}{\partial \rho_s} \tag{45}$$

is given by the classical $\sigma$ field and the derivative of the nuclear matter vertex at the local scalar density.

We are now in the position to rewrite the original Lagrangian, Eq.(8), as

$$\begin{aligned} \mathcal{L} &= \mathcal{L}^{MF} + \delta\mathcal{L} \\ \mathcal{L}^{MF} &= \mathcal{L}_B^{MF} + \mathcal{L}_M^{MF} \ . \end{aligned} \tag{46}$$

The difference between the full and the Hartree Lagrangian is contained in $\delta\mathcal{L}$. The isovector $\rho$ meson and electromagnetic interactions will be left out for the moment and we concentrate on effects from the density dependence of the vertices. Baryon mean–field dynamics are then described by the Lagrangian

$$\mathcal{L}_B^{MF} = \overline{\Psi}\left[i\gamma_\mu\left(\partial^\mu - V_\omega^\mu\left(\Gamma_\omega(\rho_0) - (\hat{\rho}_0 - \rho_0)\frac{\partial\Gamma_\omega}{\partial\rho_0}\right)\right)\right.$$
$$\left.-\left(M - \Phi\left(\Gamma_\sigma(\rho_0) - (\hat{\rho}_0 - \rho_0)\frac{\partial\Gamma_\sigma}{\partial\rho_0}\right)\right)\right]\Psi \quad . \tag{47}$$

The meson Lagrangian $\mathcal{L}_M^{MF}$ is defined as $\mathcal{L}_M$ in Eq.(8) but here only the static meson fields enter.

The conventional $\sigma$ and $\omega$ self–energies [2] are apparent in Eq.(47). The terms being proportional to the density derivatives of the vertices are the Hartree approximations to the variational derivatives of the full vertex functionals. Because they involve the normal ordered and fluctuating parts of $\hat{\rho}_0$ their expectation values vanish. In Hartree approximation these terms neither contribute to the source terms of the meson field equations, Eqs.(36) and (37), nor to the standard part $\Sigma^{(0)}$, Eqs.(40) and (41), of the baryon self–energies. By variation with respect to $\overline{\Psi}$ and taking Hartree expectation values, however, the baryon rearrangement self–energies are recovered. In total, the baryon and meson mean–field equations are retrieved from $\mathcal{L}^{MF}$ by imposing the subsidiary conditions that the baryon sources of the meson equations and the baryon self–energies are evaluated in Hartree approximation.

The difference between mean–field and full meson–baryon interactions is contained in $\delta\mathcal{L}$. Quantal mesonic and baryonic modes contribute [2, 3] but those parts will not be discussed here. In the present context contributions from the density dependence of the vertices to the residual Hartree interaction are of more interest. They are of a particular structure given by

$$\delta\mathcal{L} \simeq \Phi\left(\Gamma_\sigma(\hat{\rho}_0) - \Gamma_\sigma(\rho_0) - C(\hat{\rho}_0)\frac{\partial\Gamma_\sigma}{\partial\rho_0}\right)\overline{\Psi}\Psi$$
$$- V_\omega^\mu\left(\Gamma_\omega(\hat{\rho}_0) - \Gamma_\omega(\rho_0) - C(\hat{\rho}_0)\frac{\partial\Gamma_\omega}{\partial\rho_0}\right)\overline{\Psi}\gamma_\mu\Psi \quad . \tag{48}$$

Inserting the Taylor expansion, Eq.(33), for the vertex functionals one finds immediately that terms up to first order in the vertex derivatives are removed and in leading order $\delta\mathcal{L}$ is a quadratic form of $C(\hat{\rho}_0)$. The cancellation of the first order terms is completely due to the mean–field rearrangement terms. This indicates that rearrangement introduces additional dynamical contributions into the mean–field.

In fact, the same kind of observation was already made in non–relativistic Brueckner theory [9, 29]. As discussed by Negele [29] rearrangement self–energies describe contributions from high momentum one particle–one hole and two particle–two hole correlations to the motion of a single nucleon. They are introduced into a density dependent theory quite naturally by a fully self–consistent treatment of wave functions and interactions. In ref.[29] the corresponding diagrams are shown to be generated microscopically from the variation of the projector $\mathcal{Q}_F$ and the self–energies, respectively, which are the sources of the density dependence of a Brueckner G–matrix. Exactly this effect is taken into account in our field–theoretical formulation by applying the variational principle also to the vertices. In an effective field theory, however, it is hardly possible to identify the microscopic origin of single contributions. But, as pointed out by Negele [29], the contributions from both the Pauli–projector and the self–energies are adequately approximated by a global dependence of interactions on the density.

Summarizing this section, the mean–field equations of conventional DD theories [12, 13, 14, 15, 16] were recovered but with the important difference that baryon rearrangement self–energies are included. Their significance for an improved description of single particle properties is known from non–relativistic theory and, as far as the many–body aspects are concerned, the same arguments apply here. A diagrammatic analysis, e.g. in refs. [9, 29], shows that the rearrangement self–energies introduce high momentum polarization diagrams describing ground state correlations from the short range NN repulsion into the mean–field potential.

## 3.2 Infinite Nuclear Matter

As a first application we consider the mean–field theory of symmetric nuclear matter. In infinite matter the field equations strongly simplify and dynamical quantities like the energy–momentum tensor can be obtained in closed form. This allows to show in a particular transparent way energy–momentum conservation and the thermodynamical consisteny of the theory. Following the usual approach electromagnetic interactions are neglected. By isospin symmetry the isovector $\rho$–meson contribution vanishes identically and only the $\omega$–meson contributes to the vector potential, Eq.(21).

In the Hartree limit, Eq.(23) leads to a modified Dirac equation

$$[\gamma_\mu \tilde{k}^{*\mu} - \tilde{m}^*]\tilde{u}^*(k) = 0 \quad . \tag{49}$$

The stationary solutions are plane wave Dirac-spinors

$$\tilde{u}_r^*(k) = \sqrt{\frac{\tilde{E}^* + \tilde{m}^*}{2\tilde{m}^*}} \begin{pmatrix} 1 \\ \frac{\vec{\sigma}\cdot\tilde{\mathbf{k}}^*}{\tilde{E}^* + \tilde{m}^*} \end{pmatrix} \chi_r \tag{50}$$

where $\chi_r$ is a two–component Pauli-spinor.

The kinetic and canonical 4–momenta $\tilde{k}^{*\mu}$ and $k^\mu$, respectively, are related by

$$\tilde{k}^{*\mu} = k^\mu - \left(\Sigma^{(0)\mu} + \Sigma^{(r)\mu}\right) \quad (51)$$

and an additional shift is obtained from the rearrangement self–energies. Also the effective mass is modified

$$\tilde{m}^* = M - \left(\Sigma_s^{(0)} + \Sigma_s^{(r)}\right) \quad (52)$$

and the in–medium mass shell condition is $\tilde{k}^{*2} = \tilde{m}^{*2}$. Because of time-reversal symmetry the spacelike parts of the vector potential vanishes and $\tilde{\mathbf{k}}^* = \mathbf{k}$. The energy of the particles is given by

$$\tilde{E}^* = \tilde{k}_0^* = \sqrt{\tilde{\mathbf{k}}^{*2} + \tilde{m}^{*2}} \quad . \quad (53)$$

The scalar density is obtained in closed form as [2]

$$\rho_s = \frac{4}{(2\pi)^3} \int_{\Theta_F} d^3k \frac{\tilde{m}^*}{\tilde{E}^*} = \frac{\tilde{m}^*}{\pi^2}\left[k_F E_F - \tilde{m}^{*2} ln\left(\frac{k_F + E_F}{\tilde{m}^*}\right)\right] \quad (54)$$

with $E_F = \sqrt{k_F^2 + \tilde{m}^{*2}}$ the Fermi-energy. Integration over the momentum states inside the Fermi–sphere is indicated by $\Theta_F$. With the results of App. A the energy-momentum tensor in nuclear matter is obtained as

$$T^{\mu\nu} = \frac{4}{(2\pi)^3}\int_{\Theta_F} \frac{d^3k}{\tilde{E}^*}\left[\tilde{k}^{*\mu}k^\nu - \tilde{k}^*_\lambda \Sigma^{(r)\lambda} g^{\mu\nu}\right] + g^{\mu\nu}\left[\frac{m_\sigma^2}{2}\Phi^2 - \frac{m_\omega^2}{2}V_{\omega\lambda}V_\omega^\lambda + \rho_s\Sigma_s^{(r)}\right] \quad . \quad (55)$$

With Eq.(51), the kinetic part of $T^{\mu\nu}$ can be rewritten in terms of kinetic momenta only

$$T^{\mu\nu}_{kin} = \frac{4}{(2\pi)^3}\int_{\Theta_F} \frac{d^3k}{\tilde{E}^*}\left[\tilde{k}^{*\mu}\tilde{k}^{*\nu} + \tilde{k}^{*\mu}(\Gamma_\omega V_\omega^\nu + \Sigma^{(r)\nu}) - \tilde{k}^*_\lambda \Sigma^{(r)\lambda}g^{\mu\nu}\right] \quad . \quad (56)$$

Only the time–like components of the vector self–energies are non–vanishing in Hartree approximation and the energy density obtained from Eq.(55) is

$$\begin{aligned}\epsilon &= T^{00} \\ &= \frac{4}{(2\pi)^3}\int_{\Theta_F} d^3k\sqrt{\mathbf{k}^2 + \tilde{m}^{*2}} + \rho_B \Gamma_\omega V_\omega^0 + \frac{m_\sigma^2}{2}\Phi^2 - \frac{m_\omega^2}{2}V_{\omega\lambda}V_\omega^\lambda + \rho_s\Sigma_s^{(r)}\end{aligned} \quad .(57)$$

As shown in App.A the vector rearrangement self–energies are cancelled by compensating contributions from the baryon field equations. In the VDD case where $\Sigma_s^{(r)} = 0$ rearrangement effects are completely removed from the energy density. This is an important result for the determination of the coupling functions $\Gamma_{\sigma,\omega}$ from DB self–energies: The same energy per particle is obtained as in the DB calculation, i.e. *the equation of state remains unchanged*. The rearrangement self–energies must be included in order to obtain energy–momentum conservation, $\partial_\mu T^{\mu\nu} = 0$, as discussed in App. A

The scalar rearrangement self–energies obtained in the SDD description are only partly removed from the energy density. This is seen by expanding the square root under the integral in terms of $\tilde{m}^* - m^*$, where $m^* = M - \Sigma_s^{(0)}$ is the effective mass without rearrangement. The leading order term is cancelled but higher order terms remain as found from Eq.(57).

The pressure obtained from the energy-momentum tensor, Eq.(55), is

$$P = \frac{1}{3}\sum_{i=1}^{3} T^{ii}$$
$$= \frac{4}{3(2\pi)^3}\int_{\Theta_F}\frac{d^3k}{\sqrt{\mathbf{k}^2+\tilde{m}^{*2}}}\frac{\mathbf{k}^2}{2} - \frac{m_\sigma^2}{2}\Phi^2 + \frac{m_\omega^2}{2}V_{\omega\lambda}V_\omega^\lambda + \rho_B\Sigma^{(r)0} - \rho_s\Sigma_s^{(r)} \quad . \tag{58}$$

Different from the energy density the rearrangement potentials contribute directly to the pressure. In App. B Eq.(58) is retrieved from the thermodynamical relation

$$P = \rho_B^2 \frac{\partial}{\partial \rho_B}\left(\frac{\epsilon}{\rho_B}\right) \tag{59}$$

where Eq.(57) has to be used. Thus, the theory is thermodynamically consistent and rearrangement affects the location of the nuclear matter equilibrium point $P = 0$. Another important test for the intrinsic consistency of the model is obtained from the Hugenholtz–van–Hove Theorem [33]

$$\epsilon + P = \rho_B E(k_F) = \rho_B\left(\tilde{E}^*(k_F) + \Sigma^0\right) \tag{60}$$

which states that at equilibrium the mean binding energy per particle equals the single particle energy at the Fermi–surface. In App. B it is shown to be fulfilled only if rearrangement is included.

In a system with fixed baryon number the chemical potential is defined [1, 2] by

$$\mu = \frac{\partial}{\partial \rho_B}\epsilon$$
$$= \Gamma_\omega V_\omega^0 + \rho_B\frac{\partial \Gamma_\omega}{\partial \rho_B}V_\omega^0 + \frac{\partial}{\partial \rho_B}\left[\rho_s\Sigma_s^{(r)} + \frac{4}{(2\pi)^3}\int_{\Theta_F}d^3k\sqrt{\mathbf{k}^2+\tilde{m}^{*2}}\right] \quad . \tag{61}$$

Evaluating the term in brackets as discussed in appendix B, the chemical potential is obtained as

$$\mu = \Gamma_\omega V_\omega^0 + \Sigma^{(r)0} + \sqrt{k_F^2 + \tilde{m}^{*2}} \tag{62}$$

In the VDD description, the bare effective mass $m^*$ and $\Sigma^{(r)0}$ from Eq.(43) have to be inserted. The first two terms combine to $\Sigma^0$.

The results imply that rearrangement also contributes to infinite matter. In the first place the baryon propagators are modified. In the VDD description this can easily be taken into account by an appropriate shift of the four–momenta as in connection with Eq.(56). The situation is different in the SDD case where also the effective mass is affected by rearrangement. Eq.(62) shows that rearrangement directly affects the single particle properties in a system with fixed baryon number. Rearrangement effects should therefore be especially important in finite nuclei.

# 4 Relativistic Hartree Description of Finite Nuclei

## 4.1 Details of Numerical Calculations

The relativistic Hartree theory of finite nuclei has been discussed in very detail at many places, e.g. in refs. [2, 4, 23, 34]. Here we concentrate on the rearrangement contributions. For spherical symmetric systems as closed shell nuclei the stationary Dirac–equation is

$$H\psi = \left[ \boldsymbol{\alpha}\cdot\mathbf{p} + \beta\left(M - \Sigma_s^{(0)}(r) - \Sigma_s^{(r)}(r)\right) - \Sigma^{(0)0}(r) - \Sigma^{(r)0}(r) \right]\psi = E\psi \quad (63)$$

where only the time–like components of the vector fields contribute. The Dirac–equation is solved with the following *ansatz* for the wave function

$$\psi_{jm}^{\kappa\tau_3} = \frac{1}{r}\begin{pmatrix} F_j^{\kappa\tau_3}(r)\mathcal{Y}_{jm}^{l=j+\frac{1}{2}\omega} \\ iG_j^{\kappa\tau_3}(r)\mathcal{Y}_{jm}^{l=j-\frac{1}{2}\omega} \end{pmatrix} \quad . \quad (64)$$

The parity eigenvalue is denoted by $\kappa$ and the charge states are distinguished by $\tau_3 = \pm 1$ for protons and neutrons, respectively. The spin–angular wave function are defined by

$$\mathcal{Y}_{jm}^{l=j\pm\frac{1}{2}\omega}(\theta,\varphi) = \sum_{m_l,m_s} i^l Y_{lm_l}(\theta,\varphi)\chi(\frac{1}{2}m_s)\langle lm_l\frac{1}{2}m_s|jm\rangle \quad . \quad (65)$$

A spherical harmonics $Y_{lm_l}$ and a two–component Pauli spinor $\chi(\frac{1}{2}m_s)$ are coupled to total angular momentum $j$ with projection $m$ by a Clebsch–Gordan coefficient. The upper and the lower radial wave functions, $F(r)$ and $G(r)$, respectively, obey a system of two coupled equations [28]. By elimination of the lower component, an effective wave equation for $F(r)$ is derived

$$\left[\partial_r^2 + \frac{\partial_r(\Sigma_s + \Sigma^0)}{E + M - \Sigma_s - \Sigma^0}\left(\partial_r + \frac{\eta}{r}\right)\right.$$
$$\left. - \frac{\eta(\eta+1)}{r^2} + \left(E - \Sigma^0\right)^2 - (M - \Sigma_s)^2\right] F_j^{\kappa T_3}(r) = 0 \qquad (66)$$

where $\eta = \kappa(j + \frac{1}{2})$. $\Sigma^0$ includes isoscalar and isovector vector meson interactions and, for protons, the static Coulomb field. First and second radial derivatives are denoted by $\partial_r$ and $\partial_r^2$, respectively. The terms involving first derivatives are removed by the transformation

$$F(r) = \sqrt{E + M - \Sigma_s - \Sigma^0}\ f(r) \qquad (67)$$

and $f(r)$ is determined by a purely second order differential equation. The solutions $f(r)$ and eigenvalues are obtained with the Numerov-method [35]. The physical wave functions $F(r)$ are finally reconstructed according to Eq.(67). The lower components are obtained from

$$G(r) = \frac{1}{\sqrt{E + M - \Sigma_s - \Sigma^0}}\left(\partial_r + \frac{\eta}{r}\right) F(r) \qquad (68)$$

by numerical differentiation. The vector densities for protons and neutrons, respectively, are obtained in terms of the Dirac wave functions

$$\rho_{T_3}(r) = \frac{1}{4\pi r^2}\sum_{kj\kappa}(2j+1)\left(|F_{kj}^{\kappa T_3}|^2 + |G_{kj}^{\kappa T_3}|^2\right) \qquad (69)$$

and the scalar densities are defined accordingly by the difference of upper and lower components. Energy levels of angular momentum $j$ and parity $\kappa$ are denoted by $k$.

The densities are used as source terms in the meson field equations which are solved by representing the propagators in coordinate space [2]. Only the monopole parts are needed for spherical closed shell nuclei as considered below. The system of density dependent Hartree-equations is solved self–consistently by iteration. The iteration is started by using densities and self–energies from a full scale relativistic Thomas–Fermi calculation [5]. Typically, 15 to 20 iterations are needed until convergence of energies and wave functions of better than $10^{-4}$ is obtained.

The meson masses $m_\sigma = 550$ MeV, $m_\omega = 783$ MeV and $m_\rho = 770$ MeV are used. The $\rho$–meson coupling strenght is chosen as $g_\rho^2/4\pi = 5.19$. In ref.[15] the vertex functions $\Gamma_{\sigma,\omega}$ were determined from fits to nuclear matter DBHF self–energies. Here, the parameterizations of Haddad and Weigel [22] by second order polynomials are used. The DBHF self–energies are reproduced very accurately over a wide range of density from 0.2 $\rho_{nm}$ to 2 $\rho_{nm}$ [22]. In ref.[22] the vertices are given as functions of the baryon density which is in the spirit of the VDD parameterization. The SDD ansatz requires to re–parameterize $\Gamma_\sigma$ by the scalar density. In practice, however, this can be avoided by using that the scalar density in nuclear matter, Eq.(54), is given as a function of the baryon density through the Fermi momentum. Thus, the scalar vertex is simply given by $\Gamma_\sigma(\rho_s(k_F)) = \Gamma_\sigma(\rho_B)$. The derivative of $\Gamma_\sigma$ is determined from the relation

$$\frac{\partial \Gamma_\sigma}{\partial \rho_s} = \frac{\partial \Gamma_\sigma(\rho_s(k_F))}{\partial \rho_B}\frac{\partial \rho_B}{\partial \rho_s}$$

and

$$\frac{\partial \rho_B}{\partial \rho_s} = \left(\frac{\partial \rho_s}{\partial \rho_B}|_{\rho_B}\right)^{-1} = \frac{1}{m^*}\sqrt{\left(\frac{3}{2}\pi^2\rho_B\right)^{\frac{2}{3}} + m^{*2}}$$

which is exact in infinite matter and a good approximation in finite nuclei.

From Eq.(66), a central and a spin–orbit potential, respectively, can be extracted [34]

$$U_C = \Sigma_s^{(0)} + \Sigma_s^{(r)} - \frac{E}{M}(\Sigma^{(0)0} + \Sigma^{(r)0}) - \frac{1}{2M}\left((\Sigma_s^{(0)} + \Sigma_s^{(r)})^2 + (\Sigma^{(0)0} + \Sigma^{(r)0})^2\right) \quad (70)$$

$$U_{SO} = \frac{1}{2M}\frac{\partial_r(\Sigma_s^{(0)} + \Sigma_s^{(r)} + \Sigma^{(0)0} + \Sigma^{(r)0})}{E + M - \Sigma_s^{(0)} - \Sigma_s^{(r)} - \Sigma^{(0)0} - \Sigma^{(r)0}} \quad . \quad (71)$$

The central and the spin-orbit potentials are affected reversely by the rearrangement contributions. In leading order $U_C$ is given by the difference of the scalar and vector fields

$$U_C = \Sigma_s^{(0)} + \Sigma_s^{(r)} - \Sigma^{(0)0} - \Sigma^{(r)0} \quad (72)$$

whereas the sum of the fields enters into the spin-orbit potential. Rearrangement contributes differently to $U_C$ and $U_{SO}$ and we expect that the spin-orbit splitting is an especially sensitive measure on the VDD and SDD descriptions.

The Hartree ground state energy is obtained from the energy–momentum tensor as

$$\begin{aligned} E_{g.s.} &= \sum_{j,E_{kj}\leq\epsilon_F} (2j+1) \; E_{kj} \\ &- \frac{1}{2}\int d^3r \left[\Gamma_\sigma(r)\rho_s(r)\Phi(r) - \Gamma_\omega(r)\rho_B(r)V_\omega^0(r) - g_\rho\rho_3(r)V_\rho^0 - e\rho_P(r)A^0(r)\right] \\ &- \int d^3r \left[\rho_s(r)\Sigma_s^{(r)}(r) - \rho_B(r)\Sigma^{(r)0}(r)\right] \end{aligned} \quad (73)$$

where $E_{kj}$ are the Dirac-eigenvalues, Eq.(63), of particles in the (positive energy) Fermi-sea. $\rho_P$ is the proton density and $\rho_3 = \rho_P - \rho_N$ the isovector density.

## 4.2 Binding Energies and Density Distributions

Calculations for finite nuclei with density dependent field theories have led in the past to a fair but not overwhelmingly good description of nuclear ground state properties ( see e.g. [21, 23]). Typically, either binding energies or charge densities were reproduced reasonably well, but hardly both of them. In all cases, rearrangement was not taken into account.

The question arises whether rearrangement can improve on this situation. In Fig.3 the central and spin-orbit Dirac potentials, Eqs.(70) and (71) are displayed for $^{40}Ca$ and the Bonn A parameter set. The rearrangement contributions are seen to be repulsive. Compared to the DD case, the depth of the potentials is lowered but also a more diffuse shape is obtained. The SDD and VDD central potentials are close in strength. The differences are more pronounced in the spin-orbit potentials where the SDD and VDD potentials differ by about 50% in the surface region. Clearly, as seen below, this will lead to corresponding differences in the energy splitting of spin-orbit partners. On first sight, it could be expected that rearrangement will lower the binding energies of nuclei. However, as seen from Tab.1, rearrangement actually increases the binding. At the same time, larger charge radii are found which is in agreement with the fact that the potentials are more swallow than in a pure DD description. In order to understand the dependence of binding energies on rearrangement we have to refer to Eq.(73). The proper binding energy was obtained by subtracting the total rearrangement energy from the single particle and meson–nucleon interaction energies. Since rearrangement is repulsive this subtraction compensates the weaker binding of the single particle states and the total binding energy is increased. The same effect is found in non-relativistic HF theory with density dependent interactions [27].

The cancellation of rearrangement effects found in binding energies does not occur in one–body observables like charge density distributions and root-mean-square ($rms$) radii. They are determined directly by single particle wave functions which include the full rearrangement contribution. As can already be deferred from the potentials in Fig.3 rearrangement will generally lead to radial wave functions which are pushed out to larger radii than in a DD calculation. The repulsive effects are surviving in density distributions and, as in Tab.1, larger rms-radii are obtained. This means, that the otherwise strong correlation of nuclear binding energy and radius is partially lifted by rearrangement.

Rearrangement strongly affects the shape of density distributions. In Fig.4 charge densities from DD, SDD and VDD calculations with the Bonn A parameter set are compared to data derived from elastic electron scattering [37]. Without rearrangement (DD) the central densities and therefore the saturation properties of nuclear matter are strongly overestimated. The best description of the measured shapes and, as seen in Tab.1, also of charge radii is obtained with the VDD calculations. The theoretical point particle density distribution, Eq.(69), are folded with a gaussian proton form factor [27] with $\sqrt{\langle r^2 \rangle_p} = 0.8 \; fm$.

In Tab.1 and in Fig.5 and 6, also results for the Bonn B and C parameter sets are shown. Comparing the results of the three NN-potentials to the experimental binding energies and charge radii in Tab.2 and the measured charge distributions, the Bonn A parameters are clearly superior. The Bonn B and C potentials systematically underestimate the binding energies and give too large charge radii when rearrangement is included. Interestingly, without rearrangement the shapes and radii of the densities are described the best with the Bonn C parameter set but on the cost of a strong underestimation of binding energies (see Fig.6 and Tab.1). This observation can be understood as a consequence of the saturation properties obtained from the different parameter sets [15]. Also in nuclear matter Bonn C describes fairly well the saturation density but strongly underestimates the binding energy. With Bonn A the nuclear matter binding energy is reproduced but a too high saturation density is found. The results for Bonn B lie in between. The present DD results differ from those of Brockmann and Toki [21] although the same parameter sets are used. In part, this is related to a different treatment of isovector and Coulomb interactions and a somewhat different extrapolation of the DB results to very low densities. In ref.[21] the $\rho$ meson was not included and the Coulomb interaction is added perturbatively at the end of the calculation [36].

In ref.[29] rearrangement effects were studied microscopically in non–relativistic Brueckner Hartree-Fock calculations. Interestingly, the same behaviour as seen in Figs.4 to 6 was observed. Without rearrangement the saturation density in $^{40}Ca$ was strongly overestimated. Actually, the overestimation of the central density in Brueckner-Hartre–Fock calculations is known for many years [38]. Including rearrangement from various types of ground state correlations the calculations were found to approach systematically the measured charge density distribution (see Fig.7 of ref.[29]). In fact, the final result in [29] is in close agreement with our $^{40}Ca$ VDD calculation, Fig.4.

In N=Z nuclei isovector contributions from the $\rho$–meson are only visible through higher order Coulomb effects. Thus, they are strongly suppressed in $^{16}O$ and $^{40}Ca$. A suitable way to study the $\rho$–meson isovector mean–field is to compare the densities of different isotopes. In Fig.7 the charge densities of the symmetric $^{40}Ca$ and the asymmetric $^{48}Ca$ system are compared. In order to emphasize the isotope shift the density difference was multiplied by $r^2$. In agreement with experiment [37], the $^{48}Ca$ charge density is pushed to larger radii. This effect is coming from an intriguing interplay of isoscalar and isovector interactions. In a neutron–rich nucleus like $^{48}Ca$ the isovector $\rho$–meson mean–field is in the average attractive for protons and repulsive for neutrons. Such a behaviour would lead to a stronger binding of protons in $^{48}Ca$ as in $^{40}Ca$. However, the bulk part of the matter distribution is determined by the isoscalar interactions. For $^{48}Ca$, this means that the excess neutrons excert a static polarization on the protons. As a result, the charge density follows the total matter density such that in $^{48}Ca$ an excess proton density is found in the surface region. In Fig.7, the isotopic shift density is described very well by both the SDD and VDD rearrangement calculations. Larger deviations occur in the surface region beyond 4 to 5 fm. This indicates the limitations of the present interaction parameter sets which have been determined in infinite nuclear matter calculations. From Fig.7 it is seen, that the DD calculation without rearrangement gives only a very rough description of the isotopic shift density and, therefore, is practically ruled out.

The calculations indicate that rearrangement alters the correlations between binding energies and radii in important aspects. This observation is illustrated very clearly in Fig.8 where theoretical and experimental binding energies are shown as functions of the central charge densities of $^{16}O$, $^{40}Ca$ and $^{208}Pb$. The diagram can be considered to represent the saturation properties of finite nuclei and roughly corresponds to a nuclear equation–of–state. The experimental values arrange in a narrow band with a slope towards larger binding energies with increasing central density. This behaviour is mainly due to the properties of the $^{40,48}Ca$–isotopes. The best description is again obtained with the Bonn A parameters. The theoretical results reproduce the data surprisingly well in the VDD description. VDD calculations with the Bonn B and C are also shown in Fig.8. They range far off the data and are having the wrong slope.

## 4.3 Single Particle Energy Spectra

In single particle energy spectra more details of nuclear dynamics are seen than in global quantities like binding energies. In the present context, their properties are especially conclusive. From Eq.(49) it is evident that single particle energies include the full rearrangement contribution while they are cancelled to a large extent in the total binding energies. The potentials shown in Fig.3 indicate that further constraints on the type of density dependence might be found.

Proton and neutron single particle energies from DD, SDD and VDD calculations with the Bonn A parameter set are shown for $^{16}O$ in Tab.3, $^{40,48}Ca$ in Tabs.4 and 5 and for $^{208}Pb$ in Tab.6. When available the theoretical results are compared to experimentally observed separation energies [37]. Even if one considers the large uncertainities in the experimental values one is led to the conclusion that rearrangement improves the agreement significantly. In the average, the VDD results are again closest to the data. It should be noted that without adjustment of parameters the same quality of agreement is obtained as in non–relativistic HF–calculations with phenomenological Skyrme–interactions. Also relativistic Hartree-Fock calculations with interactions based on a parameterization of the full DBHF G-matrix [16, 23] do not lead to a better description than the present density dependent Hartree calculations with rearrangement. An exception is found in $^{48}Ca$ where the neutron energies are systematically too strong while the proton levels are described reasonably well. Very likely, these deviations are related to the iso–vector $\rho$ meson. The neutron excess of $^{48}Ca$ enhances the $\rho$ meson contributions. In $^{208}Pb$ isovector effects are less visible because they are hidden behind the stronger Coulomb interaction of the protons.

The interaction parameters originate from symmetric nuclear matter and, in fact, do not include the $\rho$ meson coupling. At present, DB calculations for asymmetric matter are not available and it is an open question how charge asymmetry would affect the coupling constants in general. A density dependence of the $\rho$ meson vertex can be expected but probably also the isoscalar couplings would be modified through higher order effects. The results of Tabs.4 to 6 lead to the conclusion that the description of the isovector channel with a density independent $\rho$ meson–nucleon coupling constant is in vain.

In Tab.7 results for the energy splitting of spin-orbit partners in the valence shells of $^{16}O$ and $^{40,48}Ca$ are shown. Here one finds, that the splitting surprisingly is described best by the DD calculations except for $^{48}Ca$. The most important contribution to the spin-orbit potential comes from the nuclear surface where the density changes rather rapidly from saturation to the free space regime. This means that the spin-orbit splitting is determined to a large extent in a density region where the DB interaction parameters cease to be reliable. The uncertainties are not coming from the parameterization of DBHF self–energies which are reproduced accurately over the whole range of densities [22]. Rather the question arises how reliable the DBHF results themselves are at low densities. Such doubts are supported by the observation that parameterizations from different DBHF calculations, e.g. [13] and [15], are in reasonable agreement for densities higher than about half of the saturation density but start to disagree drastically at lower densities.

The sensitivity of the calculations on the low density region is also indicated by the strong shift of the $1h_{11/2}$ proton level, Fig.12 and Tab.6. In the calculations the $1h_{11/2}$ state becomes the proton Fermi-level while experimentally a $3s_{1/2}$ state is observed (see Tab.6). The best result is obtained in the VDD Bonn A calculations where the two levels approach. But the $1h_{11/2}$ state remains by about 100 $keV$ above the $3s_{1/2}$ state. This behaviour is a persistent feature of many relativistic structure calculations, e.g. also in the DBHF calculations of Boersma and Malfliet [16]. Taken together with the former observations and considering the high orbital momentum of that state it is very likely that the shift is caused by a too low attraction in the nuclear surface. Interestingly, the position of the $1h_{11/2}$ proton level is reasonably well reproduced in QHD calculations with empirical coupling constants [2].

As a global test for the quality of description the mean square deviations of theoretical and experimental single particle energies were calculated. DD, SDD and VDD results for protons and neutrons in $^{208}Pb$ are shown in Tab.8. The largest $\chi^2$ values are obtained for the DD case without rearrangement. With rearrangement the results of Tab.8 are in favor of the VDD description. Comparing the Bonn A, B and C results one finds that the neutron spectrum seems to be better described in the average by the Bonn B parameter set while the proton results support the use of Bonn A. However, the differences in $\chi^2$ are insignificant and taken together with the results for binding energies and charge densities we are led to the conclusion that the Bonn A parameter is superior.

## 5 Summary and Conclusions

An effective hadron field theory with medium dependent meson–baryon vertices was presented. The dynamical structure of in–medium meson–nucleon vertices and the use of nuclear matter results in an effective quantum field theory was discussed. A density dependent field theory was obtained by a functional mapping of nuclear matter vertices onto effective meson–baryon vertices. Covariance of the field equations, energy–momentum conservation and the thermodynamical consistency of the theory were shown. The approach provides a conceptual link between a microscopic many–body description of in-medium interactions and an effective hadron field theory.

A consistent treatment of medium effects in a density dependent field theory leads to important changes in the baryon field equations. In the Euler–Lagrange equations also variations of the vertex functionals with respect to the baryon field operators have to be considered. It was shown that this gives rise to rearrangement self–energies. To a large extent their Lorentz structure is determined by the conditions that the Dirac equation has to be covariant and vertices and baryon field operators must commute. Parameterizations using the nuclear vector (VDD) and scalar densities (SDD), respectively, have been investigated. The SDD and VDD descriptions differ dynamically. With VDD vertices only vector rearrangement self–energies appear while in the SDD case both scalar and vector rearrangement self–energies were found. A mean–field Lagrangian accounting for rearrangement was derived. The relation to other approaches [21, 23] became apparent in the Hartree VDD description with vertices depending on the local baryon number density.

The essential difference to a DD theory without rearrangement lies in the dynamical structure of the mean–field. In a pure DD theory the dependence of the interactions on the bulk density is taken into account but the polarization of the background medium is neglected. In a theory with rearrangement the quasiparticles are additionally dressed by high momentum excitations of the background medium form the short range NN repulsion. A whole class of one particle–one hole and two particle–two hole diagrams is included into the mean–field such that the corresponding diagrams are cancelled in expectation values of one–body operators [9, 29]. The intermediate configurations are far off the quasiparticle energy–momentum shell.

Such virtual off-shell excitations introduce energy shifts and alter the momentum structure of wave functions but leave the quasiparticle in a stationary state. It is therefore still possible to use a Hartree or Hartree-Fock description.

Density dependent Hartree results without and with rearrangement were compared to data for charge density distributions and energy spectra of finite nuclei. From the calculations it is obvious that rearrangement significantly improves the agreement with data. Most impressively the importance of rearrangement was seen in the charge densities. Only with rearrangement a realistic description of the observed saturation densities, charge radii and binding energies could be obtained. The results indicate a sensitivity also on the low density region. Applications to light neutron–rich dripline nuclei in the neighbourhood of $^{11}Li$ are likely to provide a deeper insight into interactions at low, but non–vanishing density. The halo structure of these nuclei emphasizes contributions from the low density region [39, 40]. Different from the situation in stable nuclei the transition from the high to the low density regime is very smooth which should allow to study the intrinsic density dependence of interactions in more detail.

One has to be aware of other contributions like the coupling to low energy core excitations [10, 41] which are important for a precise description of single particle spectra. They lead to dynamical self–energies which have a particular strong influence on the valence shells. Dynamical polarization is beyond a Hartree or Hartree-Fock description. RPA methods [41] should be used but at present relativistic particle–core coupling calculations are not available.

The Hartree results favor the Bonn A parameter set and the VDD description. For this combination binding energies, charge radii and densities are well described over the whole range of the mass table as indicated by the results for $^{16}O, ^{40,48}Ca$ and $^{208}Pb$. Interestingly, also infinite nuclear matter is described the best by the Bonn A parameters [15]. In former DD Hartree calculations without rearrangement [21, 22] the Bonn C potential was found to provide a better description of finite nuclei although nuclear matter properties are poorly reproduced. Our rearrangement calculations resolve this conflicting finding and restore the dynamical consistency between nuclear matter and finite nuclei.

In the calculations, no attempt was made to optimize the input. In this sense the calculations are parameter free. The theory of section 2 could be applied equally well on an empirical basis. The density dependence of the vertices could be determined by fitting a polynomial with adjustable coefficients to data rather than to DB self–energies. In view of the lack of theoretical information on medium effects in isovector $\rho$ and other meson vertices an empirical approach could lead to valuable results on the density dependence in these interaction channels. The theory can easily be extended to hyperons and heavier baryons. Of particular interest are applications to hypernuclei and strange matter in neutron stars [5, 42] where an enhancement of medium effects can be expected.

# A  Energy-Momentum Conservation

In a density dependent hadronic field theory energy–momentum conservation is found to depend critically on the proper treatment of rearrangement. In order to avoid unnecessary complications the $\rho$-meson and the photon will be neglected. Their vertices are density independent and give standard contributions to the energy-momentum tensor which are known to be energy-momentum conserving [28].

The energy-momentum tensor is defined by

$$T^{\mu\nu} = -g^{\mu\nu}\mathcal{L} + \sum_i \frac{\partial \mathcal{L}}{\partial(\partial^\mu \phi_i)}\partial^\nu \phi_i \quad , \phi_i = \overline{\Psi}, \Psi, \Phi, V_\omega^\lambda \quad . \tag{74}$$

Inserting the Dirac equation, Eq.(23), one finds

$$\begin{aligned}T^{\mu\nu} &= i\overline{\Psi}\gamma^\mu \partial^\nu \Psi - g^{\mu\nu}\overline{\Psi}\left[\gamma_\lambda \Sigma^{(r)\lambda} - \Sigma_s^{(r)}\right]\Psi + \partial^\mu \Phi \partial^\nu \Phi + \partial^\nu V_{\omega\lambda} F^{(\omega)\lambda\mu} \\ &\quad - \frac{g^{\mu\nu}}{2}\left[\partial_\lambda \Phi \partial^\lambda \Phi - m_\sigma^2 \Phi^2 - \frac{1}{2}F^{(\omega)}_{\lambda\rho}F^{(\omega)\lambda\rho} + m_\omega^2 V_{\omega\lambda} V_\omega^\lambda\right] \quad .\end{aligned} \tag{75}$$

Energy-momentum conservation is given if $\partial_\mu T^{\mu\nu} = 0$ holds.

The divergence of the kinetic term in Eq.(75) is evaluated with the help of the Dirac equation and the corresponding equation for the adjoint baryon field

$$\left[\gamma_\mu(i\vec{\partial}^\mu - \Sigma^\mu - \Sigma^{(r)\mu}) - (M - \Sigma_s^{(0)} - \Sigma_s^{(r)})\right]\Psi = 0 \tag{76}$$

$$\overline{\Psi}\left[\gamma_\mu(i\overleftarrow{\partial}^\mu + \Sigma^{(0)\mu} + \Sigma^{(r)\mu}) + (M - \Sigma_s^{(0)} - \Sigma_s^{(r)})\right] = 0 \tag{77}$$

which gives

$$\partial_\mu(i\overline{\Psi}\gamma^\mu \partial^\nu \Psi) = \overline{\Psi}\left(\partial^\nu\left[\gamma_\mu(\Sigma^{(r)\mu} + \Sigma^{(r)\mu}) + (M - \Sigma_s^{(0)} - \Sigma_s^{(r)})\right]\right)\Psi \quad . \tag{78}$$

The divergence on the right handside of Eq.(78) acts only on the term in brackets. The derivative of the second term in Eq.(75) which includes the rearrangement potentials is split into two parts

$$\begin{aligned}&- \partial_\mu g^{\mu\nu}\overline{\Psi}\left[\gamma_\lambda \Sigma^{(r)\lambda} - \Sigma_s^{(r)}\right]\Psi = \\ &- \overline{\Psi}\left(\partial^\nu\left[\gamma_\lambda \Sigma^{(r)\lambda} - \Sigma_s^{(r)}\right]\right)\Psi - \left[(\partial^\nu \overline{\Psi}\gamma_\lambda \Psi)\Sigma^{(r)\lambda} - (\partial^\nu \overline{\Psi}\Psi)\Sigma_s^{(r)}\right] \quad .\end{aligned} \tag{79}$$

and the sum of Eq.(78) and Eq.(79) gives

$$\overline{\Psi}\gamma_\mu\Psi(\partial^\nu \Sigma^{(0)\mu}) - \overline{\Psi}\Psi(\partial^\nu \Sigma_s^{(0)}) - (\partial^\nu \overline{\Psi}\gamma_\lambda \Psi)\Sigma^{(r)\lambda} + (\partial^\nu \overline{\Psi}\Psi)\Sigma_s^{(r)} \quad . \tag{80}$$

From the first two terms in Eq.(80) one finds

$$\partial^\nu \Sigma^{(0)\mu} = \partial^\nu(\Gamma_\omega(\hat{\rho}_0)V_\omega^\mu) = \Gamma_\omega(\hat{\rho}_0)\partial^\nu V_\omega^\mu + V_\omega^\mu \frac{\partial \Gamma_\omega}{\partial \hat{\rho}_0}\hat{u}_\lambda \partial^\nu \hat{j}^\lambda \tag{81}$$

$$\partial^\nu \Sigma_s^{(0)} = \partial^\nu(\Gamma_\sigma(\hat{\rho}_s)\Phi) = \Gamma_\sigma(\hat{\rho}_s)\partial^\nu \Phi + \Phi \frac{\partial \Gamma_\sigma}{\partial \hat{\rho}_s}\partial^\nu \hat{\rho}_s \tag{82}$$

and compensating contributions to the rearrangement parts in Eq.(80) are obtained. It is apparent that such a cancellation would not have been obtained if rearrangement had been neglected in the baryon field equations. In that case, the terms involving the derivatives of the vertices would remain and lead to a violation of energy–momentum conservation, especially in finite nuclei. Eq.(81) leads to

$$\partial_\mu \left( i\overline{\Psi}\gamma^\mu \partial^\nu \Psi - g^{\mu\nu}\overline{\Psi}\left[\gamma_\lambda \Sigma^{(r)\lambda} - \Sigma_s^{(r)}\right]\Psi \right) = \Gamma_\omega(\partial^\nu V_\omega^\mu)\overline{\Psi}\gamma\Psi - \Gamma_\sigma(\partial^\nu \Phi)\overline{\Psi}\Psi \quad . \quad (83)$$

The right handside of Eq.(83) is of the same form as in the original $\sigma$-$\omega$-model. Making use of the meson field equations, Eq.(10) and Eq.(11), a straightforward calculation shows that Eq.(83) cancels with the derivative of the remaining part of $T^{\mu\nu}$ which contains the conventional meson contributions. Thus, energy-momentum conservation is fulfilled if and only if rearrangement is taken into account in the baryon field equations.

# B  Thermodynamical Consistency

Thermodynamical consistency requires the equality of the pressure obtained from the thermodynamical definition, Eq.(59), and from the energy-momentum tensor, Eq.(58), i.e.

$$\rho_B^2 \frac{\partial}{\partial \rho_B}\left(\frac{\epsilon}{\rho_B}\right) = \frac{1}{3}\sum_{i=1}^{3} T^{ii} \quad . \quad (84)$$

In the following, it is shown that this relation is fulfilled for the density dependent hadron field theory. As in Appendix A, we consider for simplicity symmetric nuclear matter and neglect the $\rho$–meson and the photon.

The left handside of Eq. (84) is evaluated with $\epsilon = T^{00}$, Eq.(57),

$$\begin{aligned} P &= \rho_B^2 \frac{\partial}{\partial \rho_B}\left(\frac{\epsilon}{\rho_B}\right) \\ &= \rho_B^2 \frac{\partial}{\partial \rho_B}\frac{4}{(2\pi)^3}\int_{\Theta_F} d^3k \tilde{E}^*(\mathbf{k}) + \frac{\partial}{\partial \rho_B}\left(\rho_s \Sigma_s^{(r)}\right) + \rho_B^2 \frac{\partial \Gamma_\omega}{\partial \rho_B} V_\omega^0 \\ &\quad - \frac{4}{(2\pi)^3}\int_{\Theta_F} d^3k \tilde{E}^*(\mathbf{k}) - \frac{m_\sigma^2}{2}\Phi^2 + \frac{m_\omega^2}{2}V_{\omega\lambda}V_\omega^\lambda - \frac{\partial}{\rho_B}\left(\rho_s \Sigma_s^{(r)}\right) \end{aligned} \quad (85)$$

where $\tilde{E}^*(\mathbf{k}) = \sqrt{\mathbf{k}^2 + \tilde{m}^{*2}}$. In spin and isospin saturated infinite nuclear matter the density is related to the Fermi-momentum by

$$\rho_B = \frac{2}{3\pi^2}k_F^3 \quad (86)$$

and thus the first term in Eq. (85) can be split into a derivative with respect to the upper boundary of the integral, i.e. to $k_F$, and a derivative with respect to the implicit density dependence of $\tilde{m}^*$ which enters via $\Gamma_\sigma$.

$$\rho_B^2 \frac{\partial}{\partial \rho_B} \frac{4}{(2\pi)^3} \int_{\Theta_F} d^3k \tilde{E}^*(\mathbf{k}) = \frac{k_F}{3} \frac{\partial}{\partial k_F} \frac{4}{(2\pi)^3} \int_{\Theta_F} d^3k \tilde{E}^*(\mathbf{k}) + \frac{4}{(2\pi)^3} \int_{\Theta_F} d^3k \frac{\tilde{m}^*}{\tilde{E}^*} \frac{\partial \tilde{m}^*}{\partial \rho_B}$$
$$= \frac{1}{3} \frac{4}{(2\pi)^3} \int_{\Theta_F} d^3k \frac{\mathbf{k}^2}{\tilde{E}^*(\mathbf{k})} + \frac{4}{(2\pi)^3} \int_{\Theta_F} d^3k \tilde{E}^*(\mathbf{k}) + \rho_s \frac{\partial \tilde{m}^*}{\partial \rho_B} \quad .(87)$$

In Eq. (87) we made use of the relation

$$\frac{1}{3} k_F^3 \tilde{E}^*(k_F) = \frac{1}{3} \int_0^{k_F} dk \frac{k^4}{\tilde{E}^*(\mathbf{k})} + \int_0^{k_F} dk k^2 \tilde{E}^*(\mathbf{k}) \quad . \tag{88}$$

Using that

$$\rho_s \frac{\partial \tilde{m}^*}{\partial \rho_B} = -\frac{\partial}{\partial \rho_B} \left( \rho_s \Sigma_s^{(r)} \right) \tag{89}$$

the equality of the thermodynamical pressure, Eq.(85), and the field–theoretical pressure, given in Eq.(58), is apparent.

Finally, from Eq.(88) the Hugenholtz–van–Hove Theorem [33] is obtained immediately,

$$\begin{aligned} \epsilon + P &= \frac{4}{(2\pi)^3} \left[ \int_{\Theta_F} d^3k \tilde{E}^*(\mathbf{k}) + \frac{1}{3} \int_{\Theta_F} d^3k \frac{\mathbf{k}^2}{\tilde{E}^*(\mathbf{k})} \right] + \rho_B \left( \Sigma_0^{(0)} + \Sigma_0^{(r)} \right) \\ &= \rho_B E(k_F) \quad . \end{aligned} \tag{90}$$

| BONN A | DD | | VDD | | SDD | |
|---|---|---|---|---|---|---|
| | $r_{ch}$ | $E/A$ | $r_{ch}$ | $E/A$ | $r_{ch}$ | $E/A$ |
| $^{16}O$ | 2.55 | 7.54 | 2.75 | 7.82 | 2.74 | 7.91 |
| $^{40}Ca$ | 3.24 | 8.55 | 3.46 | 8.79 | 3.40 | 8.95 |
| $^{48}Ca$ | 3.25 | 8.47 | 3.49 | 8.78 | 3.43 | 8.92 |
| $^{208}Pb$ | 5.14 | 7.78 | 5.48 | 8.07 | 5.34 | 8.24 |

| BONN B | DD | | VDD | | SDD | |
|---|---|---|---|---|---|---|
| | $r_{ch}$ | $E/A$ | $r_{ch}$ | $E/A$ | $r_{ch}$ | $E/A$ |
| $^{16}O$ | 2.64 | 5.59 | 2.84 | 5.81 | 2.83 | 5.89 |
| $^{40}Ca$ | 3.33 | 6.41 | 3.55 | 6.61 | 3.50 | 6.74 |
| $^{48}Ca$ | 3.34 | 6.32 | 3.58 | 6.58 | 3.52 | 6.70 |
| $^{208}Pb$ | 5.26 | 5.47 | 5.58 | 5.69 | 5.45 | 5.86 |

| BONN C | DD | | VDD | | SDD | |
|---|---|---|---|---|---|---|
| | $r_{ch}$ | $E/A$ | $r_{ch}$ | $E/A$ | $r_{ch}$ | $E/A$ |
| $^{16}O$ | 2.73 | 5.16 | 2.99 | 5.46 | 2.99 | 5.53 |
| $^{40}Ca$ | 3.44 | 5.81 | 3.74 | 6.10 | 3.70 | 6.20 |
| $^{48}Ca$ | 3.45 | 5.71 | 3.78 | 6.07 | 3.74 | 6.16 |
| $^{208}Pb$ | 5.44 | 4.72 | 5.83 | 5.01 | 5.73 | 5.14 |

Table 1: Root–mean–square charge radii $r_{ch}$ ($fm$) and binding energies per nucleon $E/A$ ($MeV/A$) of closed shell nuclei from relativistic Hartree calculations using the Bonn A,B and C parameter sets. Results without (DD) and with rearrangement in the VDD and SDD description, respectively, are shown.

| | $^{16}O$ | $^{40}Ca$ | $^{48}Ca$ | $^{208}Pb$ |
|---|---|---|---|---|
| $r_{ch}$ | 2.73 | 3.49 | 3.47 | 5.50 |
| E/A | 7.98 | 8.55 | 8.67 | 7.86 |

Table 2: Experimental values of r.m.s.–charge radii and binding energies per nucleon of the same nuclei as in table 1. Data are taken from refs. [23, 37].

<div align="center">

**$^{16}O$**

| Shell | Neutrons | | | | Protons | | | |
|---|---|---|---|---|---|---|---|---|
| | DD | VDD | SDD | exp. | DD | VDD | SDD | exp. |
| $1s_{1/2}$ | 50.6 | 40.2 | 38.6 | 47 | 46.2 | 36.0 | 34.4 | 40± 8 |
| $1p_{3/2}$ | 28.0 | 20.9 | 20.3 | 21.8 | 23.9 | 17.1 | 16.4 | 18.4 |
| $1p_{1/2}$ | 22.3 | 16.7 | 17.5 | 15.7 | 18.2 | 12.9 | 13.6 | 12.1 |

</div>

Table 3: Single particle energies of $^{16}O$ obtained with the Bonn A parameter set. Results without (DD) and with rearrangement in the VDD and SDD description, respectively, are shown. The experimental values are taken from ref. [23].

<div align="center">

**$^{40}Ca$**

| Shell | Neutrons | | | | Protons | | | |
|---|---|---|---|---|---|---|---|---|
| | DD | VDD | SDD | exp. | DD | VDD | SDD | exp. |
| $1s_{1/2}$ | 64.6 | 53.0 | 51.9 | | 55.7 | 44.8 | 43.5 | 50± 10 |
| $1p_{3/2}$ | 47.1 | 37.3 | 36.6 | | 38.6 | 29.4 | 28.5 | 34± 6 |
| $1p_{1/2}$ | 43.5 | 34.4 | 34.5 | | 35.0 | 26.6 | 26.5 | 34± 6 |
| $1d_{5/2}$ | 29.5 | 22.1 | 21.6 | 21.9 | 21.5 | 14.6 | 14.1 | 15.5 |
| $2s_{1/2}$ | 23.6 | 18.1 | 18.9 | 18.2 | 15.6 | 10.8 | 11.4 | 10.9 |
| $2d_{3/2}$ | 23.1 | 17.5 | 18.4 | 15.6 | 15.3 | 10.0 | 10.8 | 8.3 |

</div>

Table 4: Single particle energies of $^{40}Ca$ obtained with the Bonn A parameter set. Results without (DD) and with rearrangement in the VDD and SDD description, respectively, are shown.

| | | | $^{48}Ca$ | | | | | |
|---|---|---|---|---|---|---|---|---|
| | | Neutrons | | | | Protons | | |
| Shell | DD | VDD | SDD | exp. | DD | VDD | SDD | exp. |
| $1s_{1/2}$ | 65.4 | 52.8 | 51.7 | | 62.2 | 48.9 | 47.8 | 55± 9 |
| $1p_{3/2}$ | 48.3 | 38.0 | 37.0 | | 46.7 | 35.3 | 34.5 | 35± 7 |
| $1p_{1/2}$ | 45.6 | 35.9 | 35.5 | | 43.7 | 33.1 | 32.9 | 35± 7 |
| $1d_{5/2}$ | 30.9 | 23.2 | 22.6 | 16 | 29.9 | 21.3 | 22.6 | 20 |
| $1d_{3/2}$ | 25.6 | 19.2 | 19.8 | 12.4 | 24.4 | 17.2 | 17.9 | 15.3 |
| $2s_{1/2}$ | 24.6 | 19.2 | 19.9 | 12.4 | 21.7 | 15.9 | 16.7 | 15.8 |
| $1f_{7/2}$ | 14.3 | 9.5 | 9.0 | 9.9 | | | | |

Table 5: Single particle energies of $^{48}Ca$ obtained with the Bonn A parameter set. Results without (DD) and with rearrangement in the VDD and SDD description, respectively, are shown.

$$^{208}Pb$$

| | Neutrons | | | | Protons | | | |
|---|---|---|---|---|---|---|---|---|
| Shell | DD | VDD | SDD | exp. | DD | VDD | SDD | exp. |
| $1s_{1/2}$ | 72.3 | 58.9 | 58.3 | | 60.5 | 46.6 | 46.1 | |
| $1p_{3/2}$ | 65.2 | 52.5 | 51.9 | | 53.9 | 40.7 | 40.2 | |
| $1p_{1/2}$ | 64.6 | 52.0 | 51.6 | | 53.3 | 40.2 | 39.8 | |
| $1d_{5/2}$ | 56.5 | 44.9 | 44.4 | | 45.7 | 33.6 | 33.1 | |
| $1d_{3/2}$ | 55.2 | 43.9 | 43.6 | | 44.3 | 32.5 | 32.3 | |
| $2s_{1/2}$ | 51.6 | 41.4 | 41.2 | | 40.0 | 29.6 | 29.5 | |
| $1f_{7/2}$ | 46.8 | 36.6 | 36.0 | | 36.3 | 25.6 | 25.1 | |
| $1f_{5/2}$ | 44.5 | 34.8 | 34.8 | | 33.9 | 23.7 | 23.7 | |
| $2p_{3/2}$ | 39.8 | 31.5 | 31.6 | | 28.6 | 20.0 | 20.2 | |
| $2p_{1/2}$ | 38.8 | 30.7 | 31.0 | | 27.6 | 19.2 | 19.7 | |
| $1g_{9/2}$ | 36.3 | 27.7 | 27.2 | | 26.1 | 16.9 | 16.5 | 15.4 |
| $1g_{7/2}$ | 32.7 | 24.9 | 25.2 | | 22.4 | 14.1 | 14.4 | 11.4 |
| $2d_{5/2}$ | 28.2 | 21.7 | 22.1 | | 17.2 | 10.4 | 10.9 | 9.7 |
| $2d_{3/2}$ | 26.5 | 20.4 | 21.2 | | 15.6 | 9.1 | 10.0 | 8.4 |
| $3s_{1/2}$ | 25.5 | 19.6 | 20.2 | | 14.1 | 8.0 | 8.8 | 8.0 |
| $1h_{11/2}$ | 25.4 | 18.4 | 18.0 | | 15.5 | 7.9 | 7.5 | 9.4 |
| $1h_{9/2}$ | 20.5 | 14.7 | 15.3 | 10.8 | | | | |
| $2f_{7/2}$ | 17.0 | 12.2 | 12.8 | 9.7 | | | | |
| $1i_{13/2}$ | 14.5 | 9.1 | 8.6 | 9.0 | | | | |
| $2f_{5/2}$ | 14.7 | 10.3 | 11.4 | 7.9 | | | | |
| $3p_{3/2}$ | 13.6 | 9.4 | 10.4 | 8.3 | | | | |
| $3p_{1/2}$ | 12.7 | 8.7 | 9.8 | 7.4 | | | | |

Table 6: Single particle energies of $^{208}Pb$ obtained with the Bonn A parameter set. Results without (DD) and with rearrangement in the VDD and SDD description, respectively, are shown.

|  | Neutrons | | | | Protons | | | |
|---|---|---|---|---|---|---|---|---|
|  | DD | VDD | SDD | exp. | DD | VDD | SDD | exp. |
| $^{16}O$ | 5.7 | 4.2 | 2.8 | 6.1 | 5.7 | 4.2 | 2.8 | 6.3 |
| $^{40}Ca$ | 6.4 | 4.6 | 3.2 | 6.3 | 5.2 | 4.6 | 3.3 | 7.2 |
| $^{48}Ca$ | 5.3 | 4.0 | 2.8 | 3.6 | 5.5 | 4.1 | 4.7 | 4.3 |

Table 7: Spin–orbit splitting for the $1p$ shell in $^{16}O$ and the $1d$ shell in $^{40}Ca$ and $^{48}Ca$. The Bonn A parameter set is used.

$^{208}Pb$

| | Neutrons | | | Protons | | |
|---|---|---|---|---|---|---|
| Potential | DD | VDD | SDD | DD | VDD | SDD |
| Bonn A | 0.5650 | 0.0573 | 0.0966 | 0.5955 | 0.0211 | 0.0325 |
| Bonn B | 0.1472 | 0.0294 | 0.0423 | 0.1234 | 0.0910 | 0.0971 |
| Bonn C | 0.1226 | 0.0614 | 0.0593 | 0.0540 | 0.2237 | 0.2358 |

Table 8: Mean $\chi^2$-deviation of theoretical and experimental neutron and proton single particle energies in $^{208}Pb$. As indicated, results for the Bonn A, B and C parameter sets and without (DD) and with rearrangement (VDD,SDD) are shown. In all cases rearrangement significantly improves the agreement.

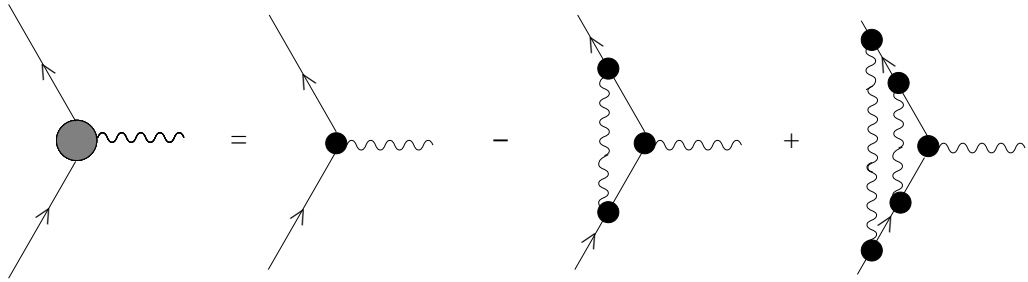

Figure 1: The diagramatic structure of meson–baryon vertices in the ladder approximation. Baryon and meson propagators are denoted by full and wavy lines, respectively. The correlated vertex (hatched circle) is obtained from the bare vertex (full circle) by an equation of Bethe–Salpeter type (see Eq.(3)). Terms up to fifth order in the bare coupling constant are shown.

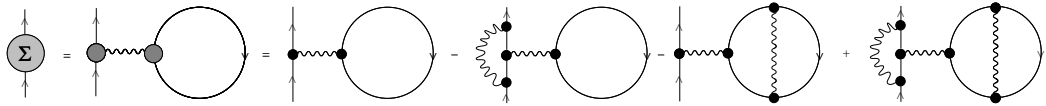

Figure 2: Diagrammatic representation of the baryon self–energy $\Sigma$ in a nuclear medium. The self–energy, Eq.(5), is displayed up to sixth order in the bare vertex (full circle). Baryon and meson propagators are denoted by full and wavy lines, respectively.

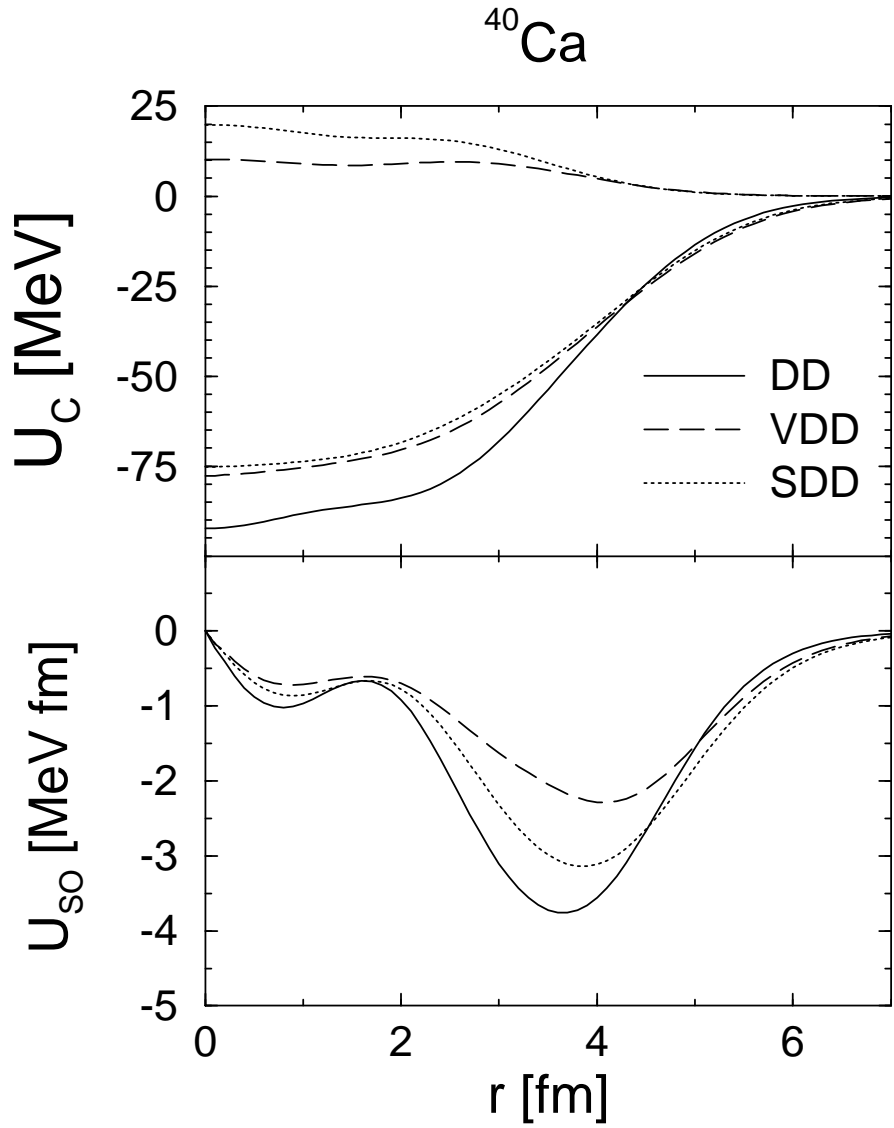

Figure 3: Dirac potentials for $^{40}Ca$. In the upper part, the central and, separately, the (repulsive) rearrangement potentials are shown for the DD, SDD and VDD description. In the lower part the corresponding spin–orbit potentials are displayed. Parameters derived from the Bonn A NN–interaction were used.

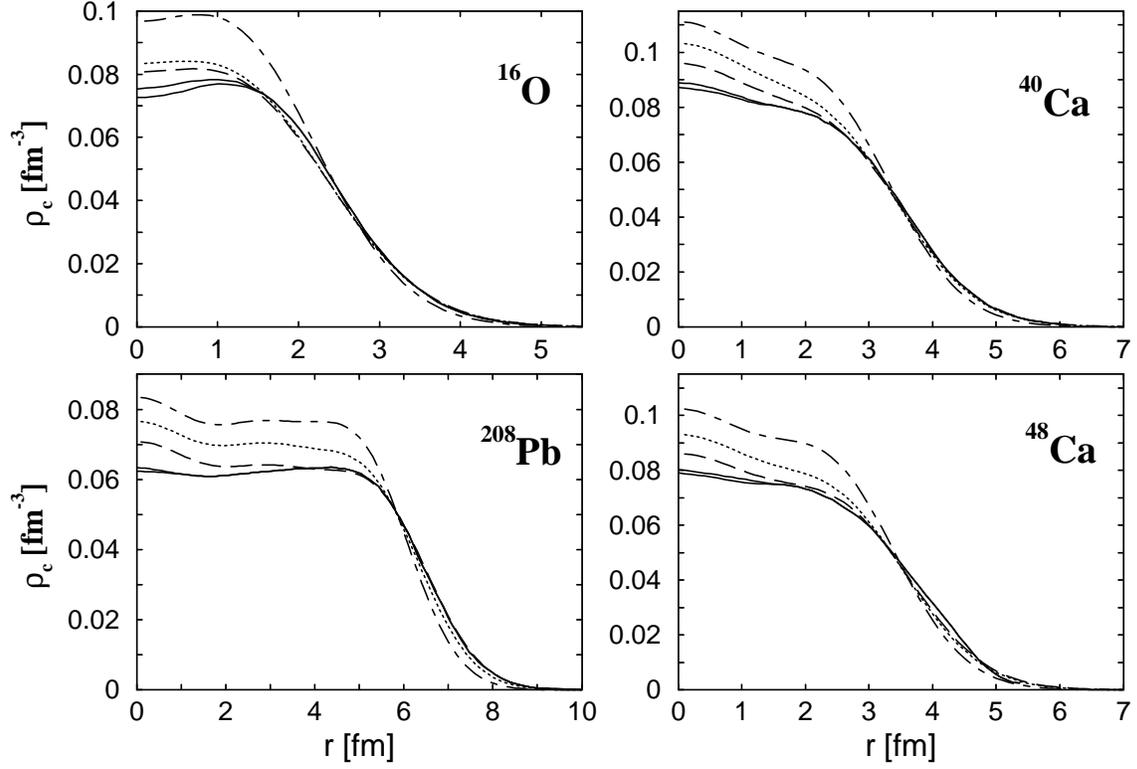

Figure 4: Charge density distributions of closed shell nuclei from relativistic density dependent Hartree calculations using the Bonn A parameter set. Results without (DD, dash–dotted) and with rearrangement in the VDD (long–dashed) and SDD (dotted) description, respectively, are shown. Experimental densities are denoted by solid lines and the uncertainities in the interior are indicated (taken from ref. [37]).

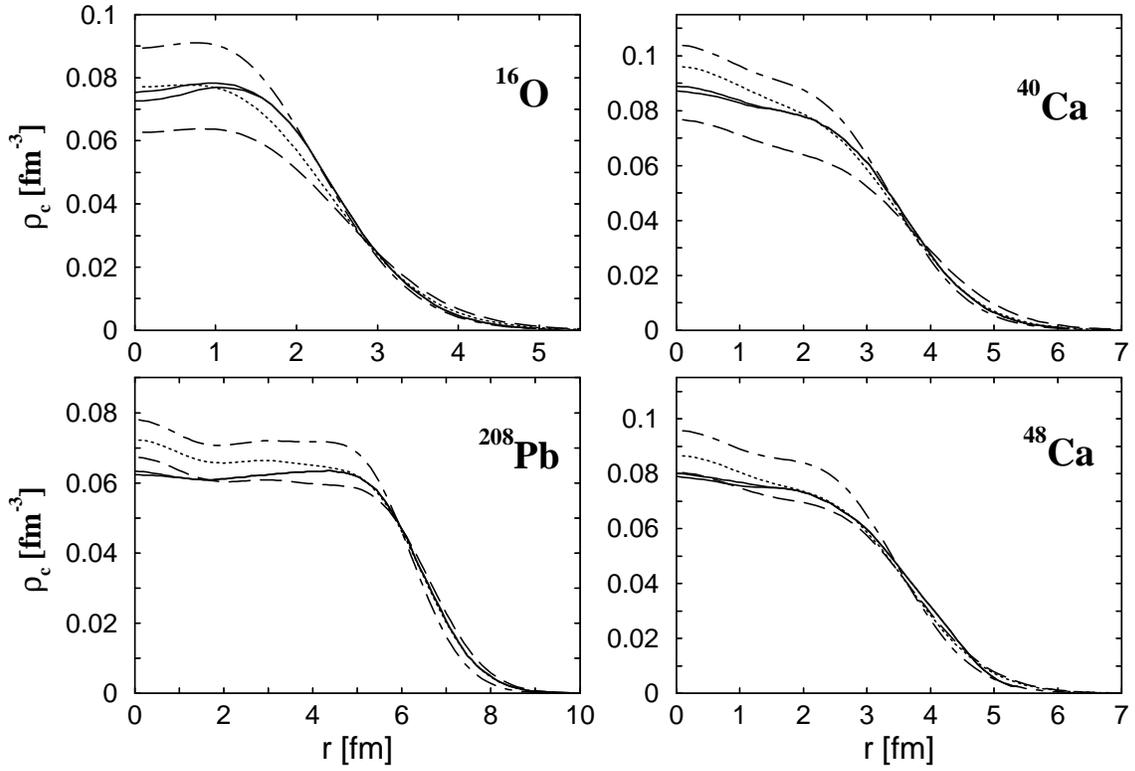

Figure 5: Charge density distributions of closed shell nuclei from relativistic density dependent Hartree calculations using the Bonn B parameter set. For captions see Fig. 4.

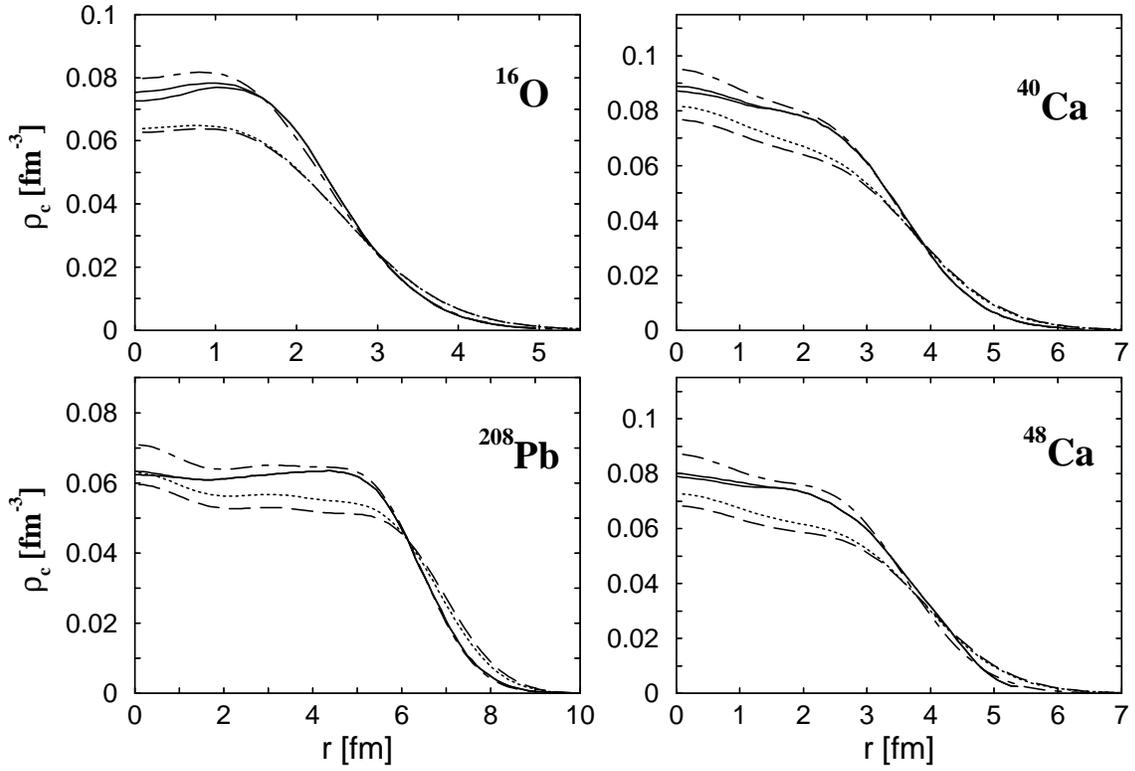

Figure 6: Charge density distributions of closed shell nuclei from relativistic density dependent Hartree calculations using the Bonn C parameter set. For captions see Fig. 4.

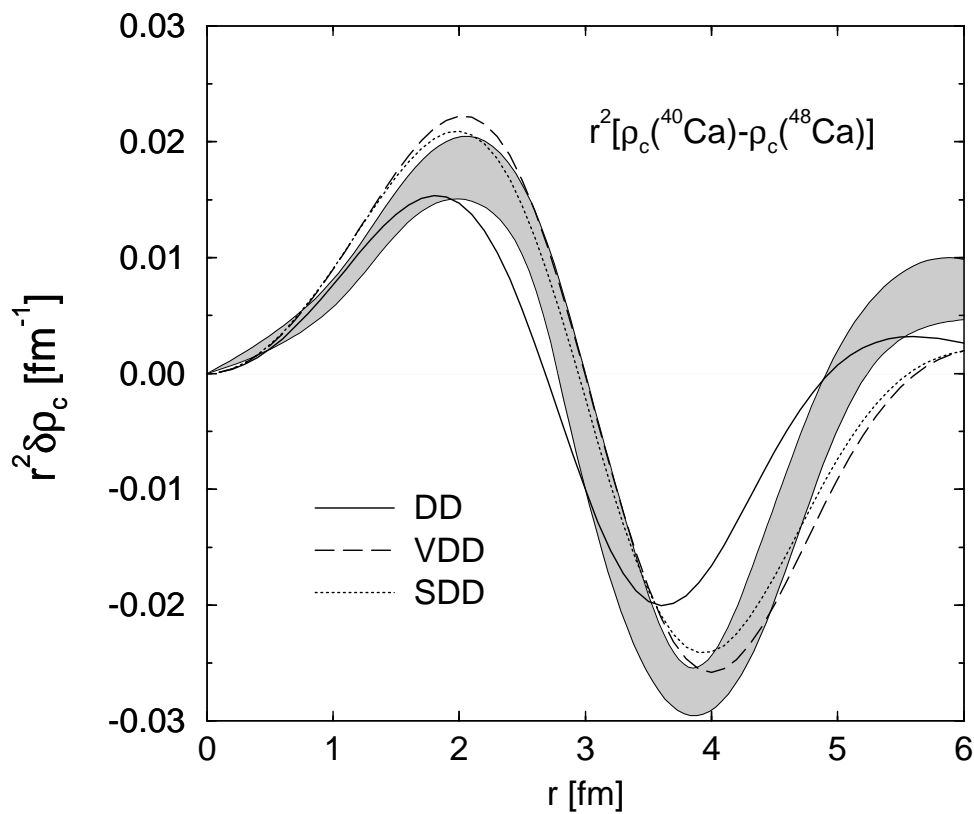

Figure 7: Isovector shift in the charge densities of $^{40}Ca$ and $^{48}Ca$. The difference of the $^{40,48}Ca$ charge densities multiplied by $r^2$ is shown for the Bonn A parameter set. Results of DD (solid), VDD (long–dashed) and SDD (dotted) calculations are compared to data [37]. Experimental uncertainties are indicated by the hatched band.

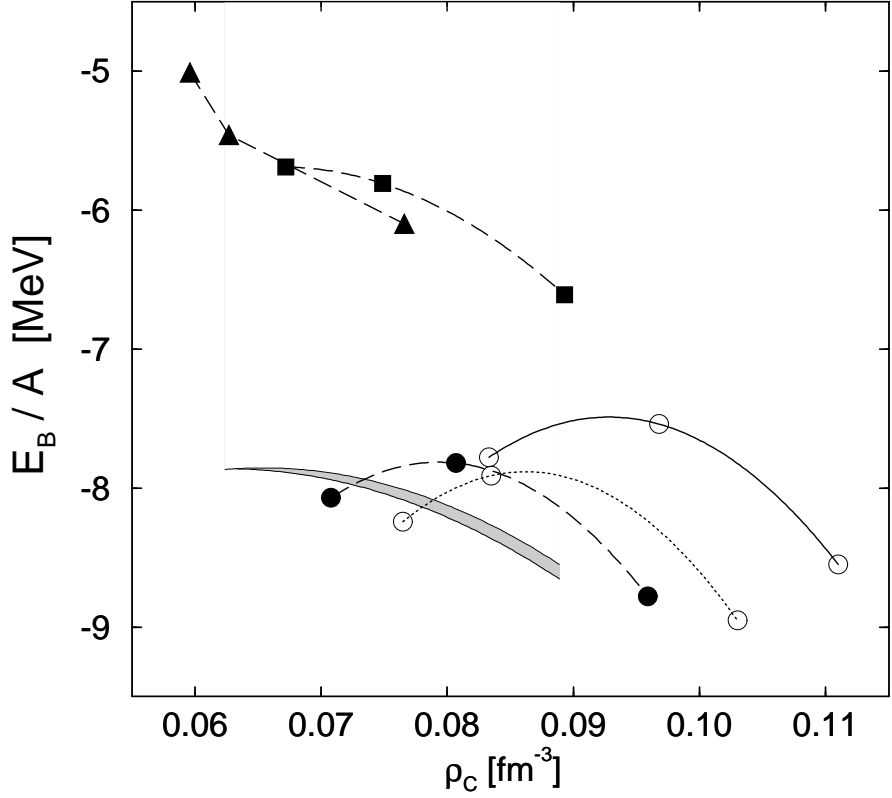

Figure 8: Equation-of-State in finite nuclei. The experimental and theoretical binding energies per particle of $^{16}O$, $^{40}Ca$ and $^{208}Pb$ are plotted against the central densities. Only for the Bonn A potential (lower set of curves) and with rearrangement the region of measured values (shaded area) is approached. The DD, SDD and VDD results are denoted by solid, dotted and dashed lines, respectively. The upper set of curves shows VDD results with the Bonn B (squares) and C (triangles) potential. The lines are drawn to guide the eye.

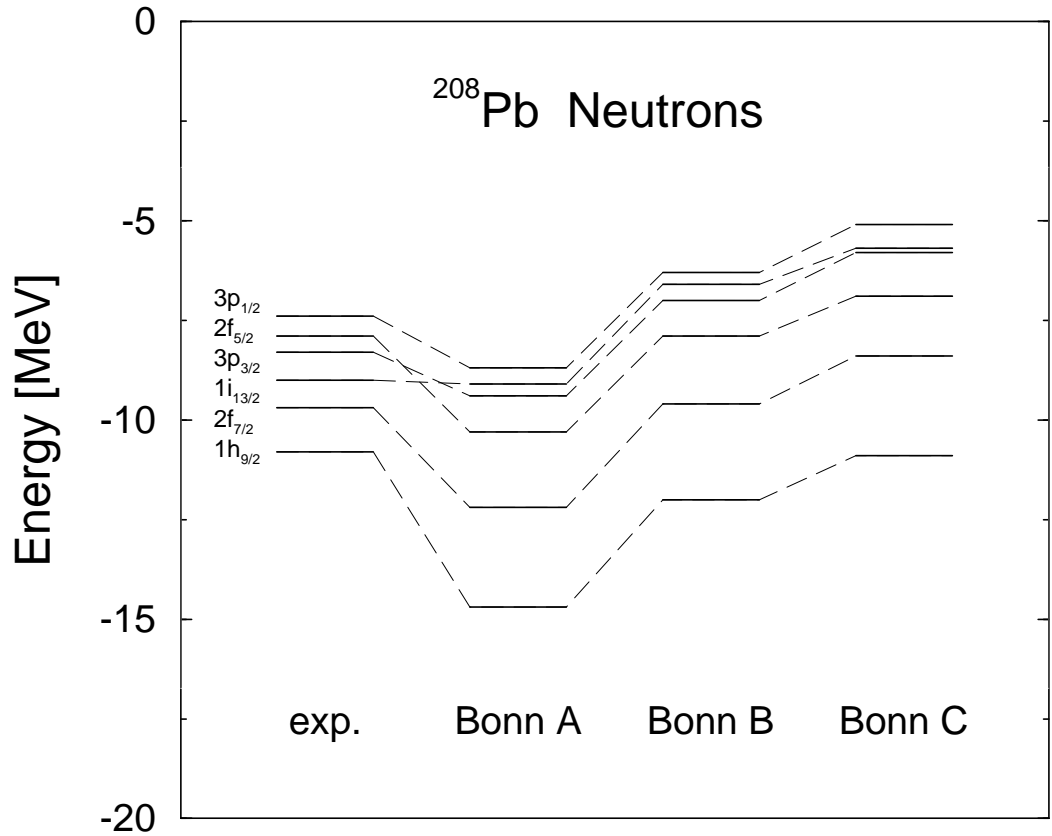

Figure 9: Neutron single particle spectrum near the Fermi energy for $^{208}Pb$ with rearrangement contributions obtained in the VDD parameterization. The calculations are performed with the Bonn A, B and C parameter sets. The experimental values are taken from ref. [37].

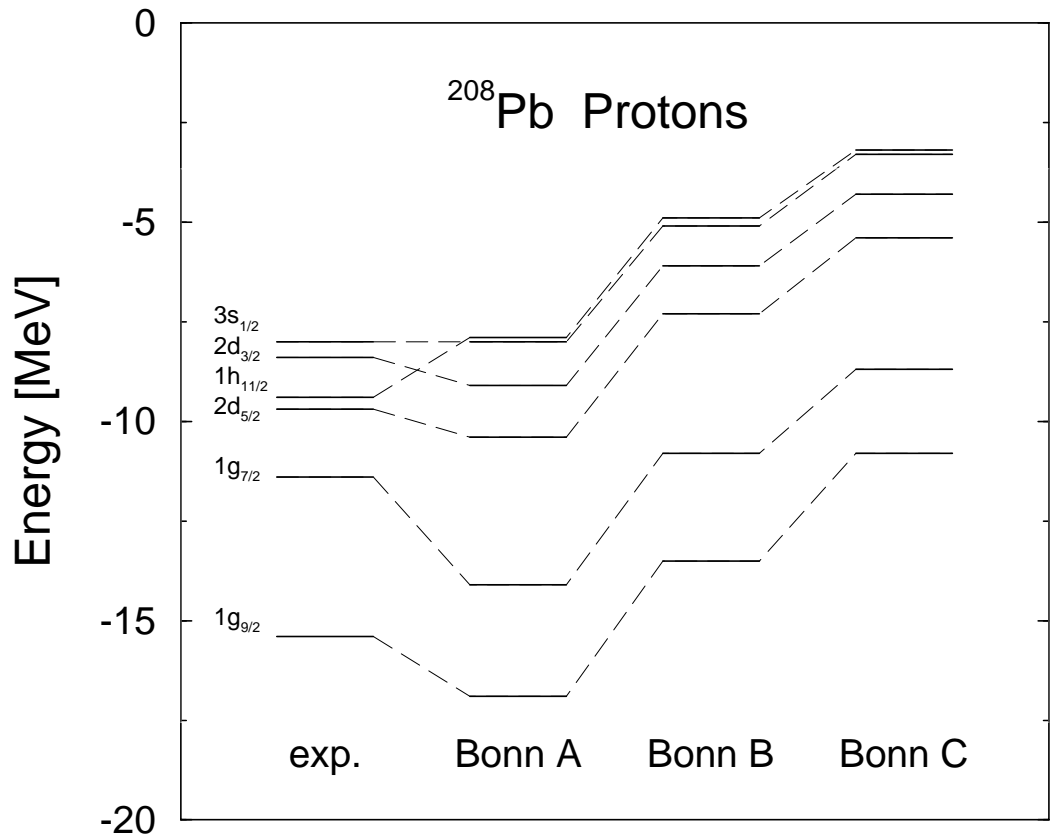

Figure 10: Proton single particle spectrum near the Fermi energy for $^{208}Pb$. For captions see Fig. 9.

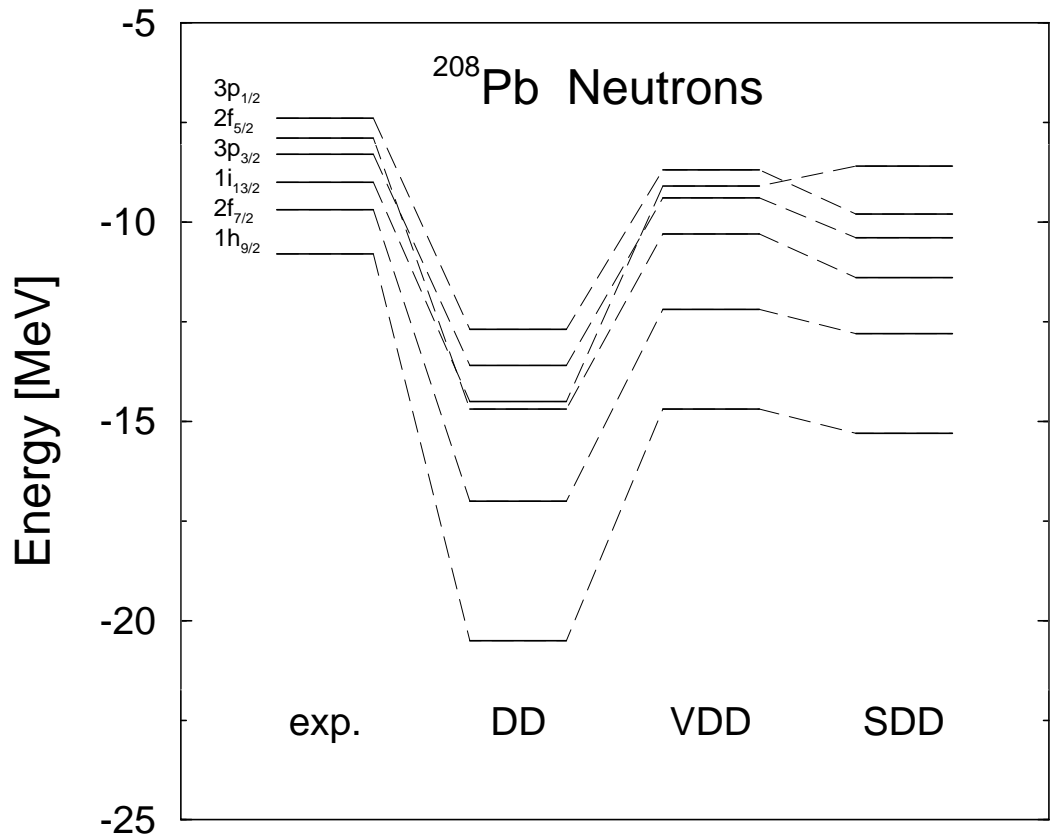

Figure 11: Neutron single particle spectrum near the Fermi energy for $^{208}Pb$ without rearrangement (DD) and including rearrangement in the VDD and SDD parametrizations. The calculations are performed with the Bonn A parameter set.

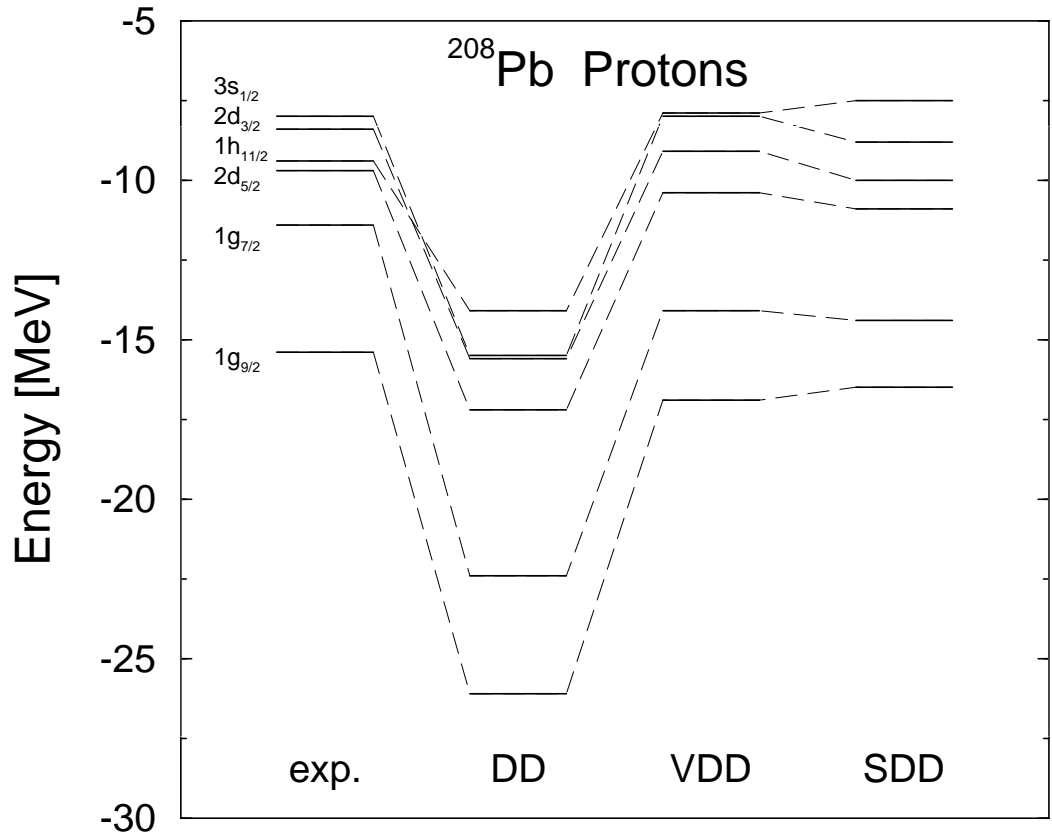

Figure 12: Proton single particle spectrum near the Fermi energy for $^{208}Pb$. For captions see Fig. 11.